\documentclass[useAMS,usenatbib]{mn2e}
\usepackage{threeparttable}
\usepackage{lscape}
\usepackage{longtable}

\usepackage{amssymb}
\usepackage{mathptmx}       % selects Times Roman as basic font
\usepackage{helvet}         % selects Helvetica as sans-serif font
\usepackage{courier}        % selects Courier as typewriter font
\usepackage{type1cm}        % activate if the above 3 fonts are
\usepackage{graphicx}        % standard LaTeX graphics tool
\usepackage{multicol}        % used for the two-column index
\usepackage[bottom]{footmisc}% places footnotes at page bottom

\usepackage{natbib}
\usepackage{aas_macros}  % Journal abreviations defined in ADS bibtex
\usepackage[normalem]{ulem}

\newcounter{subfigure}

\usepackage[usenames,dvipsnames]{xcolor}
\usepackage[total={17.8cm,24.0cm},centering]{geometry}

\title[Virgo Cluster and field dEs in 3D: I. On the variety of stellar kinematic and line-strength properties]{Virgo Cluster and field dwarf ellipticals in 3D:\\ I. On the variety of stellar kinematic and line-strength properties } 
\author[Agnieszka Ry\'{s} et al.]{Agnieszka Ry\'{s}$^{1,2}$ \thanks{E-mail:
arys@iac.es}, Jes\'{u}s Falc\'{o}n-Barroso$^{1,2}$ and Glenn van de Ven$^{3}$\\ 
$^{1}$Instituto de Astrof\'{i}sica de Canarias, 38200 La Laguna, Tenerife, Spain\\
$^{2}$Departamento de Astrof\'{i}sica, Universidad de La Laguna, 38205 La Laguna, Tenerife, Spain \\
$^{3}$Max Planck Institute for Astronomy, K\"{o}nigstuhl 17, 69117 Heidelberg, Germany
}
\begin{document}

\date{Accepted 2012 October 18. Received 2012 October 17; in original form 2012 August 2}
\date{}
% \pagerange{\pageref{firstpage}--\pageref{lastpage}} \pubyear{2011}

\maketitle

\label{firstpage}

\begin{abstract}
We present the first large-scale stellar kinematic and line-strength maps for dwarf elliptical galaxies (9 in the Virgo Cluster and 3 in the field environment) obtained with the SAURON integral-field unit. No two galaxies in our sample are alike: we see that the level of rotation is not tied to flattening (we have, e.g. round rotators and flattened nonrotators); we observe kinematic twists in 1 Virgo and 1 field object; we discover large-scale kinematically-decoupled components in 2 field galaxies; we see varying gradients in line-strength maps, from nearly flat to strongly peaked in the center. The great variety of morphological, kinematic, and stellar population parameters seen in our data points to a formation scenario in which properties are shaped stochastically. A combined effect of ram-pressure stripping and galaxy harassment is the most probable explanation. We show the need for a comprehensive analysis of kinematic, dynamical, and stellar population properties which will enable us to place dwarf ellipticals and processes that govern their evolution in the wider context of galaxy formation.
\end{abstract}

\begin{keywords}
galaxies: dwarf -- galaxies: evolution -- galaxies: formation -- galaxies: kinematics and dynamics -- galaxies: populations/stellar content
\end{keywords}

\section{Introduction}

The importance of dwarf elliptical (dE) galaxies lies in numbers. They are the numerically dominant galaxy class in clusters, by far outnumbering any other type (\citealt{sandage:1985}, \citealt{binggeli:1988}). They are low-luminosity objects (M$_B\geq - $18 mag), which makes both external influences and internal feedback mechanisms far more extreme than in more massive galaxies. Thus, they are ideal ``laboratories'' to test various mechanisms that shape galaxies. They are also a surprisingly (from a historical point of view) inhomogeneous class: which has made it challenging both to relate different dE subtypes to each other, as well as to place the whole class in the larger context of galaxy assembly and (trans)formation processes. This is perhaps best reflected by the fact that even today the astronomical community is not unanimous in their choice of the most accurate name for this galaxy class.

\subsection {Photometric, stellar population, and kinematic properties}

Much has been learned about dEs in recent years. As far as photometric studies go, the first indications for a disky nature of some dwarf early-type galaxies are found in \cite{sandage:1984} and \cite{binggeli:1991}. The first pieces of evidence for the presence of substructures such as spiral arms were found by \cite{jerjen:2000} and \cite{barazza:2002} for Virgo, \cite{graham:2003b} for Coma, and \cite{derijcke:2003} for Fornax dEs. \cite{lisker:2006a} were the first to carry out a systematic study of all Virgo dEs down to $m_B$=18\,mag covered by the Sloan Digital Sky Survey Data Release 4. Their work resulted in a new classification that divided dEs into 4 sub-types: nucleated (N), non-nucleated (nN), blue-core (bc), and disky (di) objects. They obtained the numerical contribution of each subtype to the whole class, discovered that the level of substructure scales with galaxy mass, as well as that nucleated galaxies are preferentially found in the cluster's center. Their results also suggested that a dichotomy exists in the dE group that is based on the presence of a nucleus. However, the ACS Virgo Cluster Survey \citep{cote:2006} found nuclei where they had not been seen previously, suggesting that the presumed dichotomy might have arisen from data quality limitations and that there might, in fact, exist a continuity in the nucleus-to-galaxy body ratio. With their deep photometry of 100 Virgo dwarfs, \cite{janz:2012} established that the level of substructure was previously greatly underestimated (only 25\% of their galaxies are fitted with a single S\'{e}rsic function).

Similar diversity of properties of dEs has been found through the analysis of their stellar populations. dEs display a variety of ages that seem to be uncorrelated with the luminosities of their main bodies, though \textit{nuclei} are on average younger and more metal-rich than the main bodies (Paudel 2011) and their metallicity ($Z$) does correlate with luminosity. There also seems to exist a relation between the ages and the projected distances of dEs to the Virgo Cluster's center \citep{michielsen:2008}. As for the gradients in these properties, \cite{chilingarian:2009} and \cite{koleva:2009} find that dEs show a large variety of metallicity gradients; the latter authors also note the presence of flat metallicity profiles in disky galaxies. \cite{koleva:2011} later find no relation of $Z$ gradients with luminosity, velocity dispersion, central age, or age gradient, but they do find a link between age gradients and luminosity, with fainter galaxies (M$_B\gtrsim$-17 mag) having stronger positive gradients.

Kinematic studies confirm and add to the variety. Rotating dEs are preferentially found in the cluster's outskirts and non-rotating in its center \citep{toloba:2009}. \cite{toloba:2011} find that rotationally supported dEs have rotation curves that resemble those of late-type spirals. 
The studies of \cite{geha:2002} and \cite{toloba:2011} also suggest that dEs (like the Es of \citealt{cappellari:2006} but unlike the dSphs of \citealt{wolf:2010}) are not dark-matter dominated within one effective radius.

\subsection{Relation to other galaxy classes and formation scenarios}

Determining the relation of dEs to other galaxy classes has long been a difficult task. Their seemingly simple and round appearance caused them to be classified together with giant ellipticals (hence the name) as their low-mass counterparts. A general consensus is yet to be reached regarding the (dis)continuity and the physical meaning of various scaling relations that include the dE, E, and S0 classes (e.g. \citealt{graham:2003}, \citealt{ferrarese:2006}, \citealt{kormendy:2009}). 

However, once the variety of properties of dEs became clear, the scenario in which dEs are primordial objects which expelled their gas in early stages of their evolution because of supernova explosions (e.g. \citealt{mori:1999}) could only hold for the round, pressure supported systems found in the central parts of the cluster, since they are the only ones that seem to have only old populations, with no star-formation episodes following their initial creation. The varying range of properties discussed above strongly suggests that the majority, if not all, dEs are most likely transformed late-type (spiral and irregular) galaxies that entered the cluster at later stages and had their properties shaped by external perturbations. This notion has, in fact, been around for almost three decades. The first authors to suggest it were \cite{wirth:1984} and \cite{kormendy:1985}, and it was later reiterated by \cite{binggeli:1988}, \cite{binggeli:1991}, and \cite{bender:1992}. \looseness-1

Different scenarios have been proposed in which star-forming late-type galaxies could have been transformed into the quiescent dEs we see today, the main ones being:

$(i)$ ram-pressure stripping, i.e. interactions between galaxies and the intergalactic medium (\citealt{gunn:1972}, \citealt{lin:1983}), transforms gas-rich dwarf irregulars/spirals into galaxies with very little cold gas, thus effectively stopping star formation. In this case it is expected that the intrinsic angular momentum of the progenitor should be conserved. %from geha

$(ii)$ galaxy harassment, i.e. multiple high-speed encounters with other galaxies in the cluster environment, combined with tidal heating resulting from the interactions with the cluster's potential well (\citealt{moore:1998}, \citealt{mastropietro:2005}), can remove stellar mass and significantly change the morphology, the strength of the effect likely being tied to the local environment density. In this case, a galaxy is expected to lose some (but not all) of its intrinsic angular momentum.

There exist examples of environmental transformation processes in action. We know that galaxies are falling into the Virgo Cluster and are having their gas stripped from, for example, the results of \cite{chung:2007} who show spiral galaxies with H{\sc i} tails pointing away from the cluster's center. The same authors later confirm a relation between the size and gas content of spirals as a function of clustrocetric distance \citep{chung:2009}. Indications for harassment are found in the existence of very low surface brightness dwarfs, e.g. VCC1052 of \cite{sandage:1984}, interpreted to be extreme examples of harassed objects, and in the presence of faint, round, non-rotating dEs in the central regions of the cluster (see the review in \citealt{kormendy:2012}).  \looseness-2

\begin{figure}
\centering
\includegraphics[width=0.45\textwidth]{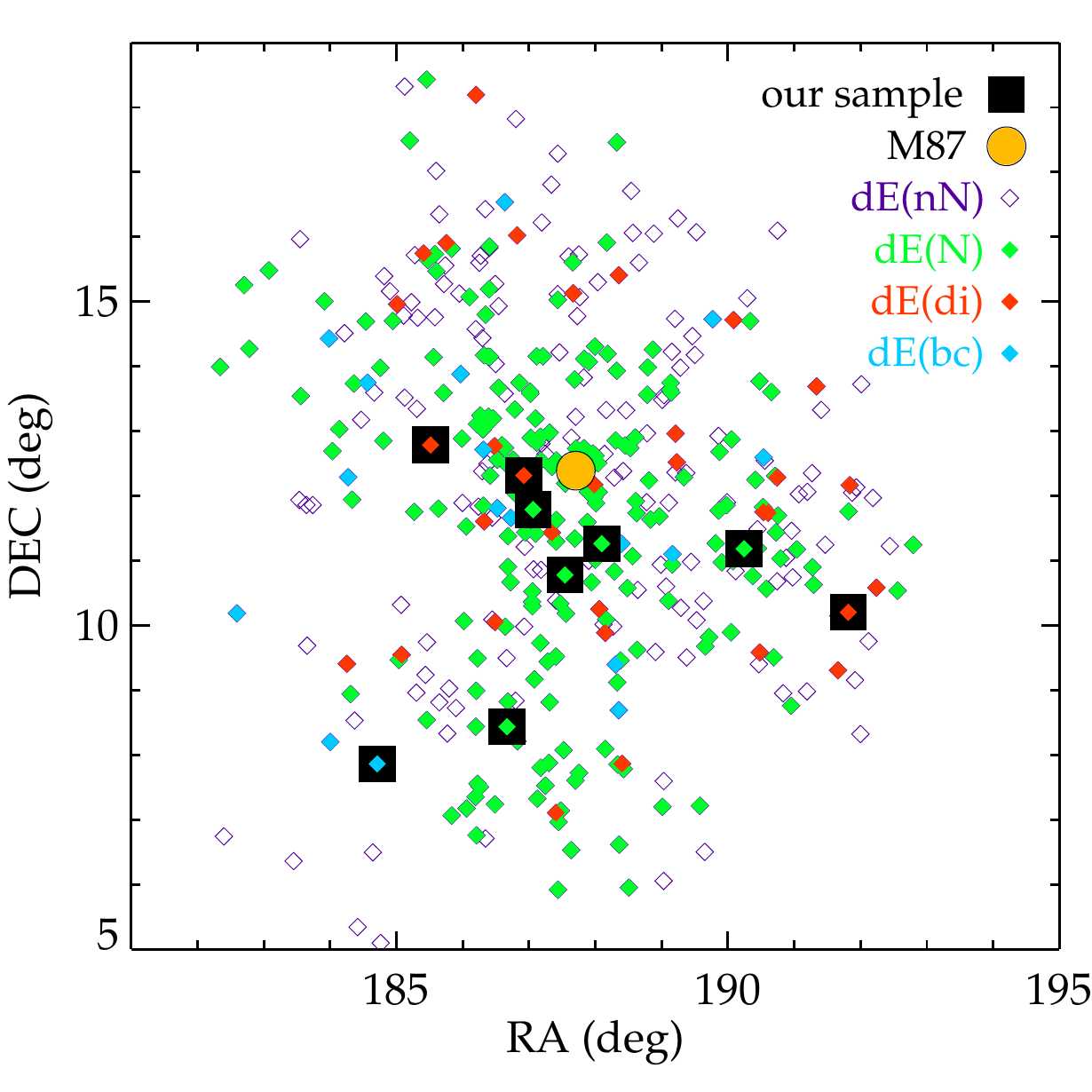}
\includegraphics[width=0.45\textwidth]{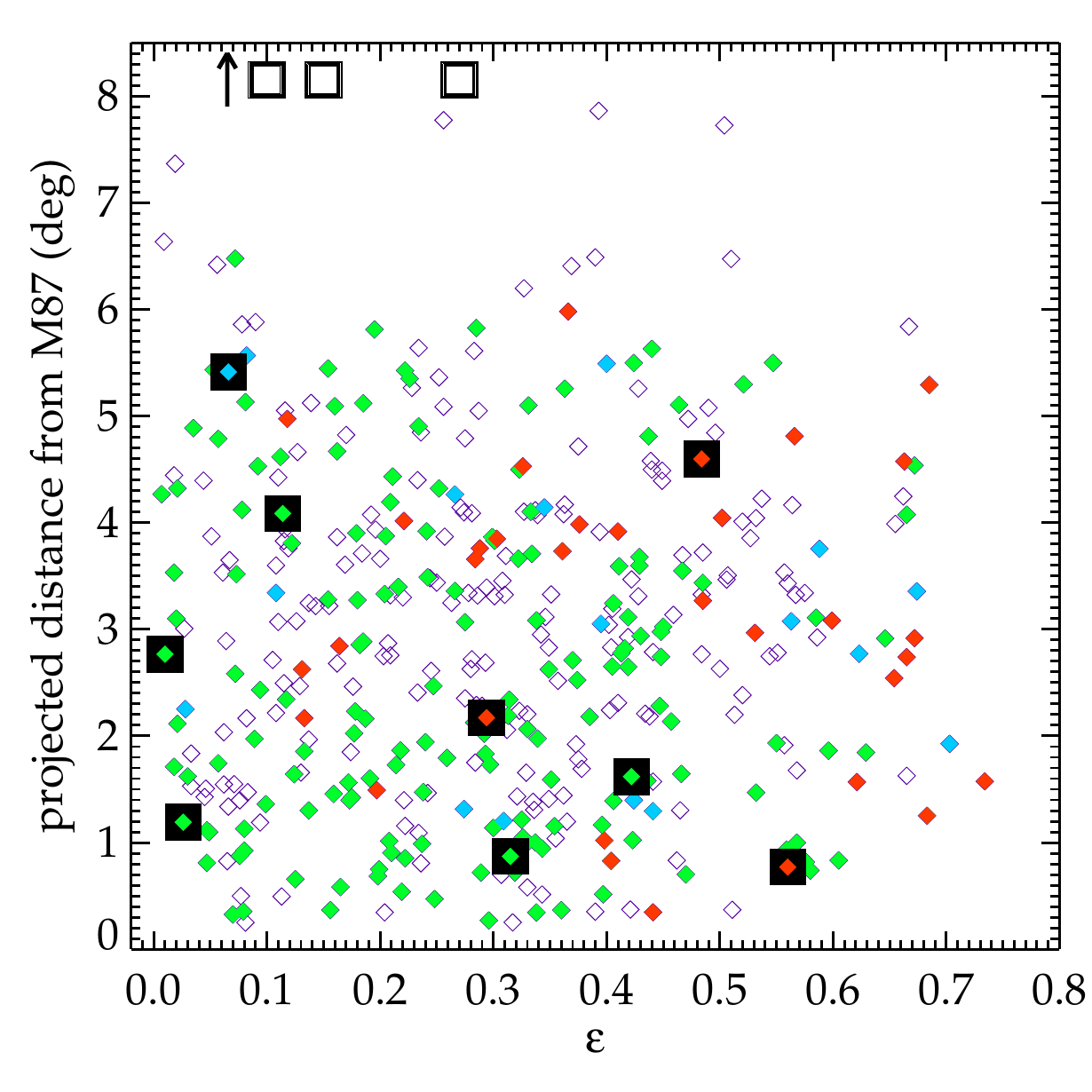}
\caption{\textit{Top panel:} positions in the cluster for all Virgo dEs from \protect\cite{lisker:2007} with the four (disky, blue core, nucleated, non-nucleated) subtypes overplotted, showing their concentration in the cluster. Our sample is shown as large filled black squares: it contains 3 dE(di)s and 1 dE(bc), the remaining 5 are dE(N)s. M87 is denoted with a filled yellow circle. \textit{Bottom panel:} projected distance from M87 vs. ellipticity for the same sample; the 3 open black squares at the top accompanied by an arrow are our field galaxies included here to show their ellipticities.}
\label{sample-paper}      
\end{figure}

The degree to which galaxies are affected by the environment depends on the galaxy mass (or, more precisely, the depth of the potential well, i.e. the degree to which it is able to gravitationally counterbalance the outside "pull") and the intensity with which is being disrupted, i.e. the environment's local density. Therefore, we can expect relations between the local density (or approximately -- clustrocentric distance) and the following quantities:
\begin{itemize}
\item \textit{Shape and the degree of substructure}. High-speed encounters with other cluster galaxies will heat up galaxies, making them rounder (\citealt{moore:1998}, \citealt{mastropietro:2005}), although the process is also suggested to be able to \textit{produce} substructures like bars or spiral arms. On the observational side \cite{lisker:2007} show that Virgo dEs containing a substructure (disks, bars, or spiral arms) are not centrally concentrated, but the dEs without substructures \textit{are}. We also know from \cite{sabatini:2005} and \cite{derijcke:2010} that the late-to-early type dwarf ratio decreases  with decreasing clustrocentric distance.
\item \textit{Degree of ordered over random motion}. As gas is (ram-pressure) stripped from a galaxy, the system is heated up and so the $V/\sigma$ drops. Still, it is harassment that is expected to be the primary process responsible for removing large portions of rotation, as predicted by \cite{moore:1996} and  \cite{mastropietro:2005} and observationally suggested by \cite{toloba:2009}.
\item \textit{Dark to stellar matter ratio}. The simulations of \cite{moore:1998} and \cite{smith:2010} suggest that (at a fixed radius) dark matter might be lost more easily due to its different orbital distribution. No observational confirmation of this trend has thus far been available.
\item \textit{Global stellar population properties}. Various environment-dependent mechanisms, such as ram-pressure stripping \citep{boselli:2008} or strangulation \citep{boschf:2008}, are said to be able to quench star formation, though their relative importance is not yet known. A number of studies find negative age and/or positive metallicity global (i.e. \textit{among}, not \textit{within} galaxies) gradients: \cite{tortora:2011}, \cite{ellingson:2001}, \cite{smith:2009}, and \cite{weinmann:2009} find that environment affects colors/star-formation rates; this results from the fact that central galaxies have resided longer in the cluster and have thus experienced environmental influences for a more extended period of time.
\end{itemize}

\subsection{This study}

We can see that a substantial amount of effort has recently been put into trying to understand the dE class. What is still missing is an ``all-inclusive'', comprehensive analysis of \textit{all} observable properties that is later matched with simulations in order to put qualitative constraints on the (trans)formation processes.

\begin{table*}
\caption{(1) Galaxy names as in the Virgo Cluster Catalog (VCC) of \protect\cite{binggeli:1985}, New General Catalog (NGC), or as a GALEX identification number (ID), (2,3) J2000 coordinates (from \textit{NASA Extragalactic Database (NED)}, http://ned.ipac.caltech.edu/), (4) for Virgo galaxies: projected distances from the cluster's center \protect\citep{lisker:2007}, (5) surface brightness fluctuation (SBF) or globular cluster luminosity function (GCLF) distances  for Virgo galaxies where available (\protect\citealt{mei:2007}$\mathrm{^{1}}$, \protect\citealt{jerjen:2004}$\mathrm{^{2}}$ or \protect\citealt{jordan:2007}$\mathrm{^{3}}$) otherwise an average Virgo distance assumed$\mathrm{^{4}}$, for NGC\,3073 a SBF distance from \protect\cite{tonry:2001}, the distances to ID\,0650 and ID\,0918 were calculated using their radial velocities, all distances are corrected to H0 = 73 km/s/Mpc for consistency with \protect\cite{mei:2007}, (6) morphological types (\protect\citealt{lisker:2007} for the Virgo objects, \protect\citealt{michielsen:2008} / \textit{NED} database for the field galaxies), (7) ellipticities at 1 effective radius from \protect\cite{lisker:2007}, (8) r-band effective radii determined by fitting aperture photometry profiles on SDSS images with R$^{1/n}$ growth curves, as detailed in \protect\cite{falcon:2011b}, (9) r-band apparent Vega magnitudes, (10) position angles of our observations, (11) and total on-source exposure times.} 
 \begin{threeparttable}
\centering
% \vspace{0.5cm}
\begin{tabular}{|p{1.3cm}|r|r|r|r|r|r|r|r|r|r|}
\hline
object&$\mathrm{\alpha}$           &$\mathrm{\delta}$          &$\mathrm{R_{M87}}$ & distance & type & $\mathrm{\epsilon}$ &R$\mathrm{_e}$ & m$_r$ & PA  & t$_{\mathrm{exp}}$\\ 
      &$(hh$ $mm$ $ss.ss)$&$(dd$ $mm$ $ss.s)$&($^{\mathrm{o}}$) &  $(Mpc)$ &      &           &$('')$&$(mag)$&$(deg)$& $(h)$             \\ 
(1)   &(2)                &(3)               &(4)                &(5)       &(6)   &(7)        &(8)   &(9)    &(10) &(11)\\
\hline
VCC\,0308 & 12 18 50.90 & +07 51 43.4 &5.42&16.50$\mathrm{^{4}}$&dE(di;bc)&0.07&18.7 $\pm$ 1.3&13.32 $\pm$ 0.08& 73.4&5.0\\
VCC\,0523 & 12:22:04.13 & +12:47:15.1 &2.21&16.50$\mathrm{^{4}}$&dE(di)   &0.29&27.9 $\pm$ 3.4&12.60 $\pm$ 0.10&320.0&5.0\\
VCC\,0929 & 12 26 40.50 & +08 26 08.6 &4.09&14.86$\mathrm{^{2}}$&dE(N)    &0.11&22.1 $\pm$ 3.4&12.65 $\pm$ 0.12&  3.3&5.0\\
VCC\,1036 & 12 27 41.24 & +12 18 57.2 &0.78&16.07$\mathrm{^{2}}$&dE(di)   &0.56&17.2 $\pm$ 1.5&13.13 $\pm$ 0.07&112.1&5.0\\
VCC\,1087 & 12:28:17.88 & +11:47:23.7 &0.88&16.67$\mathrm{^{1}}$&dE(N)    &0.31&28.6 $\pm$ 3.1&12.85 $\pm$ 0.10&107.0&4.0\\
VCC\,1261 & 12:30:10.35 & +10:46:46.3 &1.62&18.11$\mathrm{^{1}}$&dE(N)    &0.42&19.7 $\pm$ 2.0&12.87 $\pm$ 0.09&313.0&5.0\\
VCC\,1431 & 12 32 23.41 & +11 15 46.9 &1.19&16.14$\mathrm{^{1}}$&dE(N)    &0.03& 9.6 $\pm$ 0.5&13.60 $\pm$ 0.06&155.4&5.0\\
VCC\,1861 & 12:40:58.60 & +11:11:04.1 &2.79&16.14$\mathrm{^{1}}$&dE(N)    &0.01&20.1 $\pm$ 2.6&13.41 $\pm$ 0.12&122.0&4.5\\
VCC\,2048 & 12 47 15.29 & +10 12 12.8 &4.63&14.45$\mathrm{^{3}}$&dE(di)   &0.48&16.5 $\pm$ 2.1&13.08 $\pm$ 0.09& 17.9&5.0\\
% \hline
NGC\,3073 & 10:00:52.08 & +55:37:07.7 & -  &32.8~               &dE/S0  &0.15&16.1 $\pm$ 2.1&12.98 $\pm$ 0.09&120.0&5.5\\
ID\,0650$^{\dag}$& 14 04 15.96 & +04 06 43.9 & -  &25.9~                &dE/S0  &0.10&20.1 $\pm$ 3.3&13.73 $\pm$ 0.14&145.3&6.0\\
ID\,0918$^{\ddag}$& 14 58 48.72 & +02 01 24.7 & -  &16.3~                &dE/E   &0.27& 6.4 $\pm$ 1.2&13.79 $\pm$ 0.12& 12.1&5.0\\
\hline
\end{tabular}
\label{observations} 
\textit{Notes}: $^{\dag}$\,other name: UGC\,08986, $^{\ddag}$\,other name: CGCG\,020-039.
\end{threeparttable}
\end{table*}

In response to these issues we have started a study of dEs in the Virgo Cluster using the SAURON integral field unit (IFU). Using integral-field spectroscopy enables us to simultaneously obtain spectra at each position in the field of view and study both kinematics and stellar populations of the galaxies in two dimensions, giving an unprecedented view of the mass distribution and star formation histories. The two-dimensional capabilities and large spatial coverage of SAURON are key to clearly identifying any substructure and robustly determining any radial trend in these galaxies. These goals cannot be easily achieved with traditional long-slit spectrographs, where in order to achieve comparable spatial coverage multiple pointings would be required.

This is the first of a series of papers in which we investigate the properties of dEs on a sample of 12 galaxies, 9 in the Virgo Cluster and 3 in the field. The paper is structured as follows. Our sample selection, observations, and data reduction are presented in Section~2. In Sections~3 and 4 we describe the methods used in the analysis of the kinematic and line-strength properties of our sample. Section~5 presents our results, and the discussion and conclusions are in Section~6. In the other papers of this series we will construct dynamical models of dEs, infer their detailed star formation and enrichment histories, and compare the global relations of dEs to other galaxy classes. All of this together will help us understand and provide more insight on the relations listed in the previous subsection. \looseness -1

\section[Data]{Data}
\subsection[Sample selection]{Sample selection}
\label{sec:sample}

The Virgo Cluster is the closest large concentration of dE galaxies beyond the Local Group and presents itself as an excellent environment in which to study these galaxies. In our target selection we used the catalog of \cite{lisker:2007} to select our cluster targets and the catalog of the MAGPOP-ITP collaboration (see \citealt{michielsen:2008}) to select our field galaxies.

As for our cluster objects, for our first observing run we selected the brightest dwarf galaxies from the Virgo luminosity function, attempting to pick objects for which long-slit data showed hints of peculiarities. Later, we decided to add objects in such a way as to probe galaxies at different values of ellipticity and distance from Virgo's central galaxy M87 (see Figure~\ref{sample-paper}). This was motivated by the assumption that the extent of the environmental influence is broadly a function of clustrocentric distance. Thus, our objects are in the 0.01-0.56 ellipticity range and 0.78--5.42$^{\mathrm{o}}$ (0.22--1.56\,Mpc) projected distance range. \looseness-2

The limitations of our attempts at achieving this particular parameter coverage are obvious. Projection effects influence both the observed flattening as well as the perceived location in the cluster. For example, one of our targets,VCC\,1261, appears to lie close to the center ($\sim$\,1.6$^{\mathrm{o}}$ in projection), but we later discovered from the distance measurements of \cite{mei:2007} that it is, in fact, located $\sim$1.6 Mpc \textit{behind} M87, which translates into their intrinsic separation of $\sim$5.8$^{\mathrm{o}}$ at the average Virgo distance. \looseness -1

Thus, any discussion about clustering properties needs to take the \textit{depth} of the cluster into account. For example, objects belonging to subclasses for which no \textit{projected} clustering is found (dE(di)s of \citealt{lisker:2007}, see top panel of Figure~\ref{sample-paper}) are in fact \textit{intrinsically} preferentially found in the cluster's outskirts. Also, getting individual intrinsic axial ratios requires prior information on the dynamical structure of galaxies (though statistical properties of whole subclasses can be assessed, see, e.g. \citealt{lisker:2007}). All this means that not only the sample size, but also the difficulty in determining intrinsic characteristics contribute to the fact that our sample cannot be regarded as representative. 

The reason for the inclusion of field galaxies in our sample was to enable a comparative analysis of cluster vs. field environments. We first preselected objects with which we were able to fill the parts of our nights when Virgo could not be observed. From this set we picked those with the highest surface brightnesses. We also tried to select galaxies for which the closest neighbors were found at different projected distances, so that we have in our sample both NGC\,3073 which is likely strongly affected by its large companion galaxy, as well as ID\,0650 for which no neighbor is found within a 0.5$^{\mathrm{o}}$ projected radius. Their ellipticities are in the 0.10--0.27 range.  No kinematic analysis has thus far been available for 2 out of 3 of the targets (ID\,0650 and ID\,0918). 

In total, the chosen galaxies have surface brightnesses at the edge of the SAURON field that are not lower than $\mu_V\approx24.0$\,mag/arcsec$^2$. They are nucleated and non-nucleated objects, half of which were known to host a disk/have a substructure. All but two have additional spectroscopic (long-slit) and photometric data available. The positions, observing details, and photometric properties for each observed object are listed in Table~\ref{observations}.

\subsection{Observations}

\begin{figure*}
\centering
\includegraphics[width=1.00\textwidth]{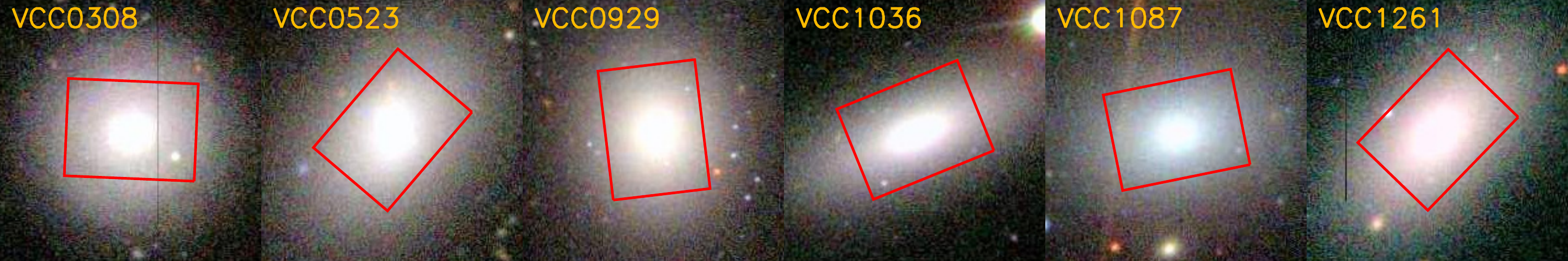}
\includegraphics[width=1.00\textwidth]{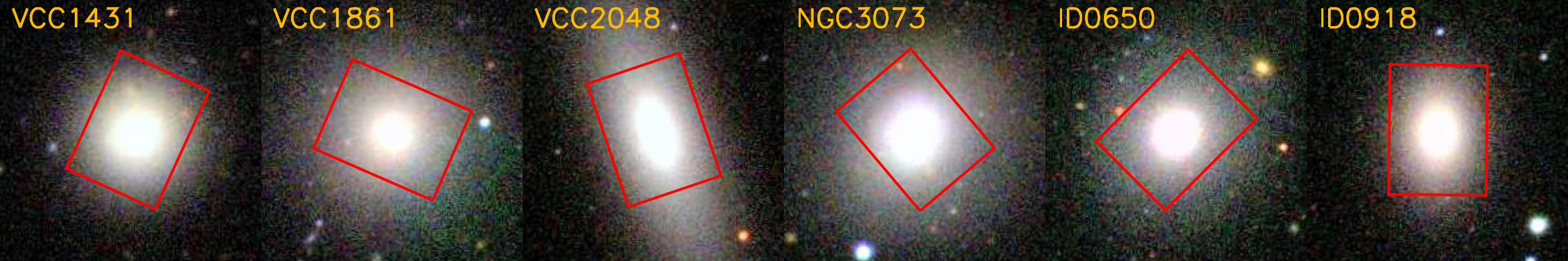}
\caption{SDSS color images (80x80'', or approximately 6.4x6.4\,kpc) with the SAURON FoV overplotted. North is up and East left.}
\label{all-pointings}      
\end{figure*}

The observations were carried out over 8 nights\footnote{out of a total of 12 that had been awarded to this project -- we lost 3 nights in Apr 2010 and 1 in Apr 2011 due to adverse weather conditions.} (3 in January 2010 and 5 in April 2011) with the SAURON integral field unit mounted on the 4.2m William Herschel Telescope at the Roque de los Muchachos Observatory, Spain. We were able to cover the galaxies out to about one effective radius (R$\mathrm{_e}$). The footprints of all the pointings overlaid on SDSS images are shown in Figure~\ref{all-pointings}.

SAURON has a wavelength range of 4760-5300\,\AA. It includes a set of potential emission lines (H$\beta$,[OIII],[NI]) and stellar absorption lines (H$\beta$, Fe5015, and Mg$b$) which can be used to investigate stellar populations. We used SAURON's low resolution mode, which gives a field of view of 33x41\,arcsec$^2$, spectral resolution (FWHM) of ~3.9\,\AA~and spatial sampling of 0.94x0.94\,arcsec$^2$ per each of the 1431 object lenslets. 

Each galaxy was typically exposed for 10x1800\,s. The exceptions are 2 field galaxies where 0.5/1.0\,h was added due to their slightly lower surface brightnesses, and 2 Virgo targets where 0.5/1.0\,h was lost due to problems with rising cryostat temperature during our first run. Small offsets were introduced among the pointings in order to improve sampling as well as deal with the bad columns in the CCD. A dedicated set of 146 sky lenslets, pointing 1.9 arcmin from the main field, was used to perform accurate sky  subtraction. As our objects have on average surface brightness a few magnitudes below the sky level at the edge of the field, we have imaged a number of blank fields (1 per night on average) to track sky variability and optimize sky subtraction. To allow for good calibration of line-strength indices we observed a number of stars during each run, covering a broad range of spectral types.  Specifically, we included stars from the MILES library \citep{sanchez-blazquez:2006} in order to calibrate our measurements. We also observed spectrophotometric standard stars each night in order to calibrate the response function of the SAURON system.

\subsection{Reduction}

We followed the procedures described in \cite{bacon:2001} for the extraction and calibration of the data using the specifically designed XSAURON software developed at the Centre de Recherche Astrophysique de Lyon (CRAL).  

First, the raw frames were preprocessed in a standard way which typically includes overscan, bias and dark subtraction. Since the dark current values for the CCD used (EEV13) are less than 1.0~e$^{-}$/pixel/h, its effect on the exposures is negligible.

The next step was to create an extraction mask, i.e. a model that relates each pixel on the CCD to its corresponding lenslet and wavelength. The resulting table, specifying lenslets' geometrical positions on the CCD and their profiles, was used to extract spectra from all preprocessed frames and create raw datacubes ($\alpha$, $\delta$, $\lambda$). The cubes were then wavelength calibrated using neon (arc) lamp exposures, which contain 11 useful emission lines in the SAURON wavelength range. These neon exposures, taken before and after every target frame, were used to estimate the possible small offsets between the science frame and the extraction mask. This was done using a cross-correlation function between its two arc exposures and the arc exposure used for the mask creation.

Flatfielding was a two-step procedure. The regular flatfield frames were created one for each observing night using twilight exposures (spatial flatfield) and continuum (tungsten lamp) exposures (spectral flatfield). Since our galaxies fall a few orders of magnitude below the sky level at the edge of the field, an additional flatfielding step was included. We took several 900s ``blank field'' exposures by pointing the telescope at relatively empty points in the sky. We then created a median of the frames from different days to improve the signal-to-noise ratio. The resulting datacube was smoothed at each wavelength (i.e. each monochromatic frame) and then divided by this wavelength's median to obtain a normalized cube. Because we could not exclude night-to-night variability, we decided to check how combining certain subsets of sky frames into a superflat and/or scaling the gradient in the combined product would affect our science frames. Such an adjustment was needed for our Run 2 objects since we opted to use Run 1 skyframes due to their better quality. Once the optimal superflat was obtained, we multiplied our regular flat fields by the superflat, and these modified flats were incorporated back into our pipeline. Dividing the cubes by the superflat removed the remaining flat-field residuals. 

\begin{figure*}
2\centering
\includegraphics[width=1.0\textwidth]{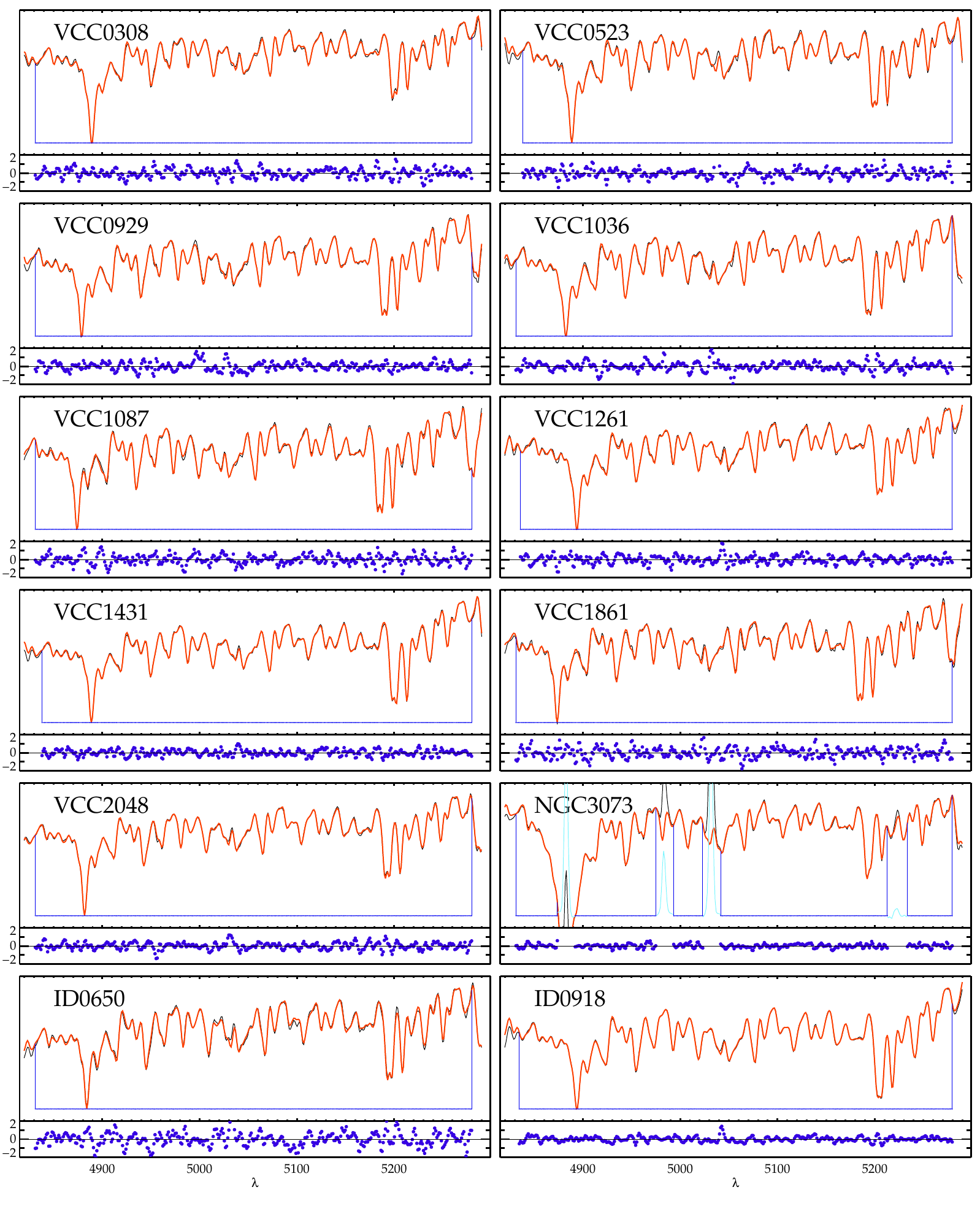}
\caption{Example pPXF fits to the galaxy spectra collapsed within 3 arcsec to illustrate the high quality of the data and the accuracy of the fits. Galaxy spectra are shown in black and the fitted models in red. The dark blue lines show the regions used for fitting. The ordinate values are plotted on normalized scales. The S/N of these spectra exceeds 100 per pixel. For NGC\,3073, emission-line regions were excluded from the fits. In the small panels below each galaxy spectrum we plot the residuals between the data and the model, shown as a percentage of the original value (with the y-axes ranges of $\pm$2\,\%).}
\label{ppxf-fit} 
\end{figure*}

Cosmic rays were removed by filtering out statistically impossible values using a wavelength direction filter run on spectra divided by the median of a few spatial neighbors.

Spectral resolution varies slightly within the field of view, with the highest values in the center and lowest in the outskirts of the field. We therefore measured spectral resolution of each lens by comparing our twilight exposures with a high-resolution solar reference spectrum and then homogenized it to a common value which was less than or equal to the value obtained for 95\% of the lenses. This was done through applying a gaussian corresponding to the difference of the squares of the actual and target resolutions. This meant using a value worse than the actual one for some parts of the field but it is a necessary step, performed to avoid irregular broadening of lines when one needs to combine images of an object located at different positions on the chip. We adopted the value of FWHM=3.89\,\AA~for both runs.

We performed sky subtraction by computing a median value of the 146 dedicated sky lenslets and then subtracting that value from the object spectra. The superflatfielding procedure described above helped to optimize this step since the sky lenslets spectra (located at the edge of the CCD) were the ones most affected by any initially present flatfield irregularities.

Our spectrophotometric standards were used to perform flux calibration. A flux correction curve was created to correct for a different instrumental response for different wavelengths. This was done by comparing the observed spectrum of a flux star with that of a reference spectrophotometric spectrum and then applying the resulting curve to all science frames (see also \citealt{kuntschner:2006} for more details).

We checked the accuracy of our flatfielding process by letting our twilight frames go through the same reduction steps as the galaxy frames (with the exception of sky subtraction since the twilights themselves are used for that purpose) and then building histograms of the values across the individual frames, which were expected to be essentially flat. Indeed, the variations for all tested files are always within 1\%, and within 0.5\% on average.

To merge the fully calibrated cubes we truncated them to a common wavelength range. Then, the cubes' spatial coordinates were recentered and the fluxes of corresponding lenslets were co-added using a median filter. The resulting cubes were in the process resampled to a spatial scale of 0.8x0.8\,arcsec$^2$.

\section{Stellar kinematics}
\subsection{Stellar kinematic measurements}
\label{kininfo}

In order to accurately measure mean stellar velocity $V$ and velocity dispersion $\sigma$, a certain minimum signal-to-noise (S/N) level is required. To achieve this, we spatially binned our data using the Voronoi two-dimensional binning algorithm developed by \cite{cappellari:2003}. In this method one starts from the central bin (with the highest S/N) and accretes closest neighboring lenslets until the required S/N has been reached, then starting a new bin. We imposed a minimum S/N of 30 for the kinematic measurements presented here. Each time, before binning data data, we also deselected those spatial pixels (spaxels) for which the S/N$_{spaxel}$ was lower than the set minimum (S/N$_{spaxel}$ is, effectively, a combination of local surface brightness and the number of frames that fall on a given spaxel). This extra step was needed since for objects as faint as ours, high S/N is critical to recovering accurate kinematic parameters values. Simply adding up the signal of individual spaxels to form larger bins will lead to lower overall S/N if some  of the input spaxels have low S/N, and, in our case, showed as a spread of $\sigma$ values that exceeded the nominal measurement errors. Thus, we wanted to know what the minimum S/N$_{spaxel}$ was that would still allow us to recover accurate kinematic measurements. After trying out a few values, we adopted S/N$_{spaxel}$=7, which roughly corresponds to surface brightness of $\mu_V\approx23.5$\,mag at the edge of the fields (slightly depending on the object), as a minimum acceptable value.  By filtering out low-S/N spaxels we made sure that our final binned spectra were not contaminated by low-quality measurements.

In the present study, stellar absorption-line kinematics were derived for each galaxy by directly fitting the spectra in the pixel space using the penalized pixel-fitting method (pPXF) of \cite{cappellari:2004}. The method fits a stellar template spectrum convolved with a LOSVD to the observed galaxy spectrum in pixel space (logarithmically binned in wavelength). The algorithm finds the best fit to the galaxy spectrum by employing nonlinear least-squares optimization, and returns the mean velocity $V$, velocity dispersion $\sigma$, and, if possible, the higher order Gauss-Hermite moments $h_3$ and $h_4$, which describe the asymmetric and symmetric departures of the LOSVD from a simple gaussian, respectively \citep{marel:1993}. % from %AMW 3.3  * Paper III, 2.5.1

The procedure of creating the optimal combined stellar template spectrum was repeated for each bin using stellar spectra from the updated MILES library \citep{falcon:2011a} containing spectra of 985 stars at 2.50\,\AA~FWHM. From these we selected a subsample of $\sim$150 stars in such a way that we still had a uniform parameter coverage.

Error estimates were obtained through Monte-Carlo simulations. For each binned spectrum we ran a loop of 100 realizations, each time creating a simulated input spectrum by taking the original spectrum and perturbing it using a spectrum of random values drawn from a range determined by the difference between the original data and the best-fit model. \color{black}{} Once we had the 100 realizations, we built histograms and took the mode with the larger of the error bars (since they were asymmetric) as the final computed error value for each bin. 

Example fits to the spectra of central regions ($\le$ 1.5 arcsec) of all objects are shown in Figure~\ref{ppxf-fit}. The galaxy spectra are shown together with their best-fit models and the data-model residual plots to illustrate their excellent agreement and the high quality of the data. \looseness-2

\begin{figure*}%[!t]
\centering
\includegraphics[width=0.99\textwidth]{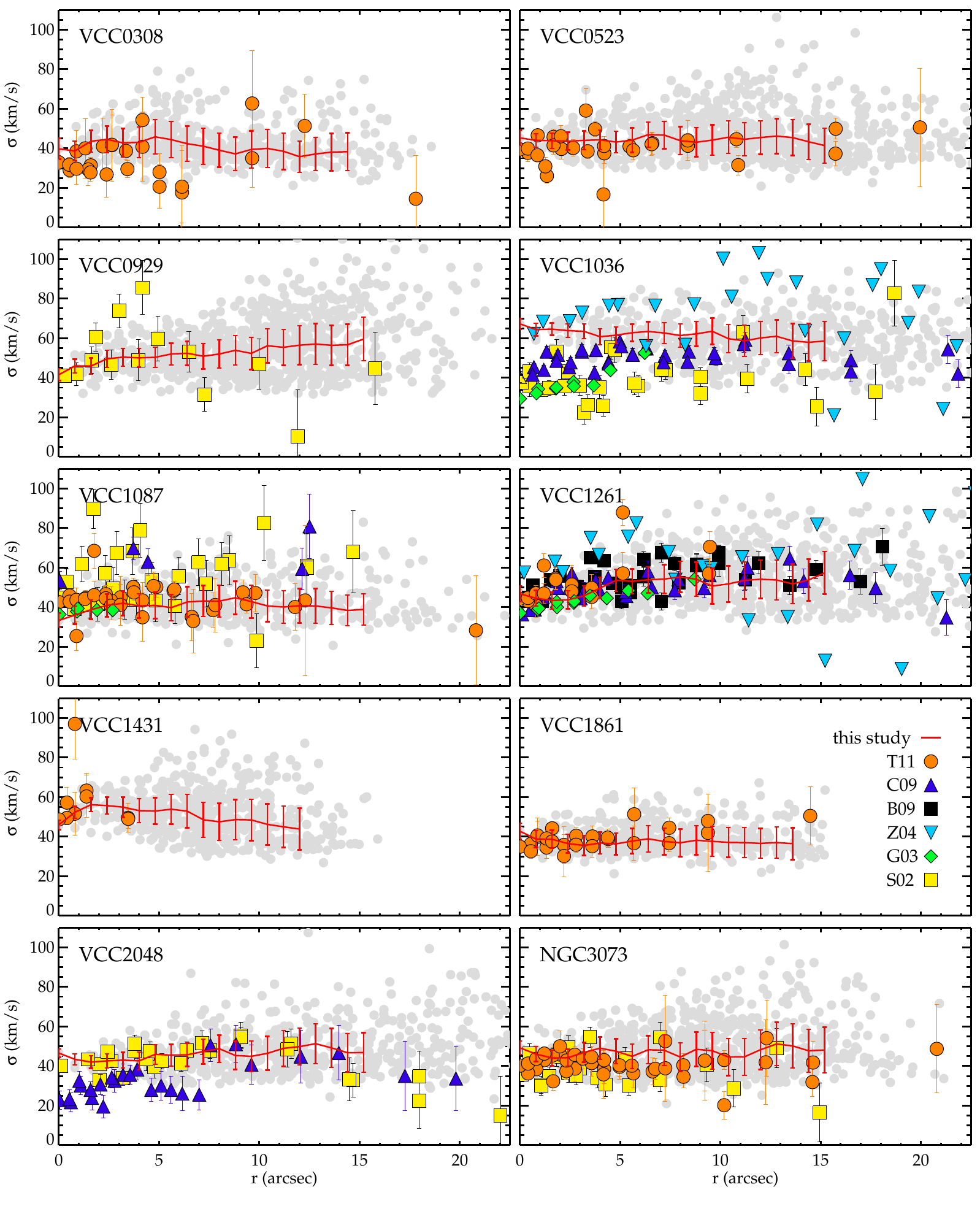}
\caption{Comparison of the published $\sigma$ profiles with the ones obtained in this study. The light-gray circles show all our bins and in red we show profiles obtained by averaging the bin values over ellipses, the overplotted errors are average errors for bins at a given elliptical distance. Literature values are shown using different colors/symbols as denoted in the VCC1861 panel. The following abbreviations were used: 1) \protect\cite{pedraz:2002}: P02, 2) \protect\cite{simien:2002}: S02, 3) \protect\cite{geha:2003}: G03, 4) \protect\cite{vanzee:2004}: Z04, 5) \protect\cite{beasley:2009}: B09, 6) \protect\cite{chilingarian:2009}: C09, and 7) \protect\cite{toloba:2011}: T11. For VCC\,0929, VCC\,1036, and VCC\,1261 some of the published profiles extend beyond what is presented in the figure.}
\label{profile_comparison} 
\end{figure*}

\begin{figure}
\centering
\includegraphics[width=0.6\columnwidth,angle=270]{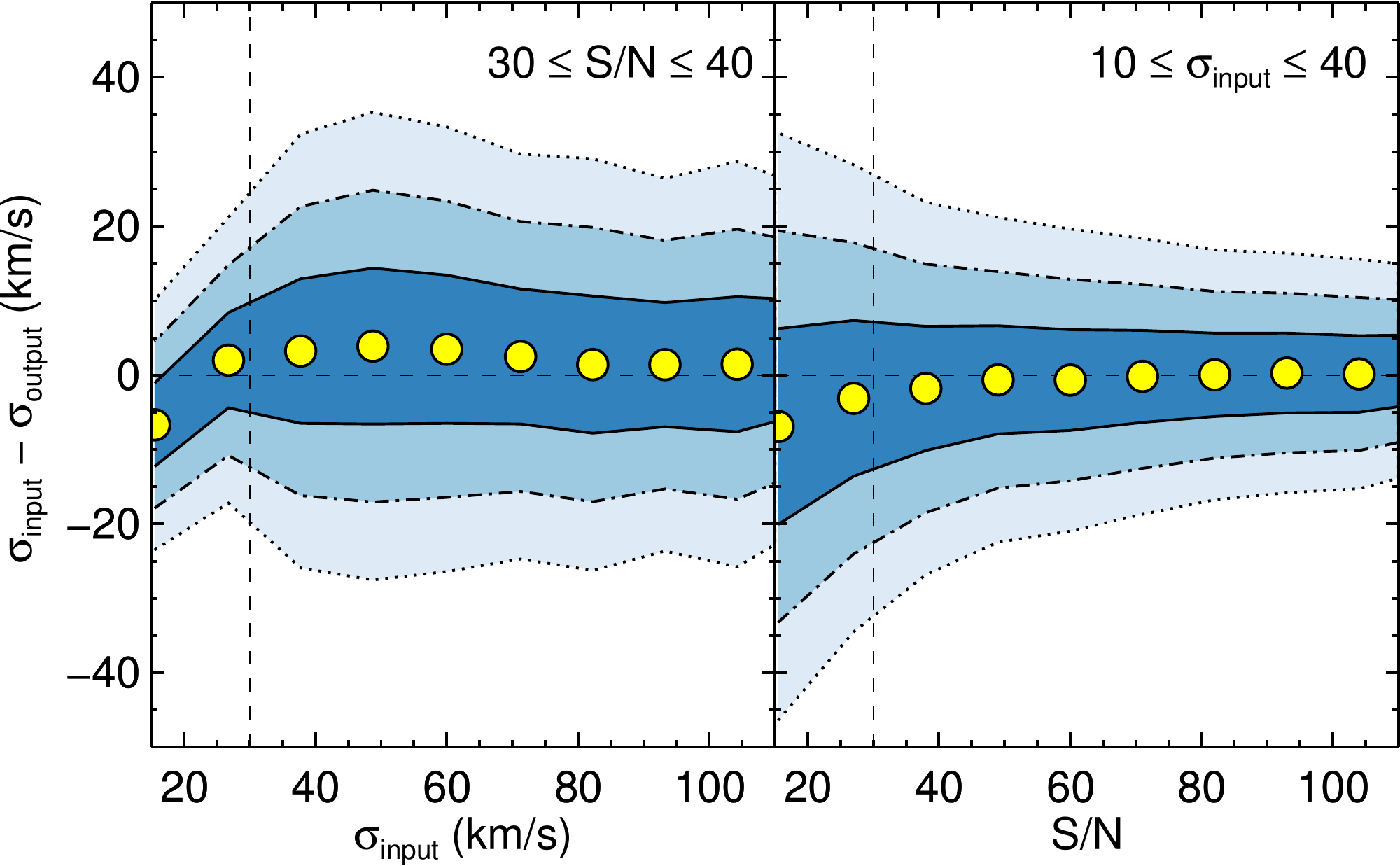}
\caption{Stellar velocity dispersion recovery tests. \textit{Left panel}: the difference between the input and recovered $\sigma$ is plotted against the input value. \textit{Right panel}: the same difference is now plotted against the signal-to-noise of the input spectrum. In both plots the different shades of blue correspond to 1, 2, and 3\,$\sigma$ levels of certainty and the dotted vertical lines show the minimum signal-to-noise level, S/N=30, of our bins. The tests in both panels only consider the lowest $\sigma$ and S/N regimes of our data. As shown in Figures~\ref{ppxf-fit} and \ref{profile_comparison}, the quality of our data and the comparison with the literature are generally much better.}
\label{sigma-sims}      
\end{figure}

\subsection{Stellar velocity dispersion \& comparison with literature}

\begin{figure*}
\centering
\includegraphics[width=1.00\textwidth]{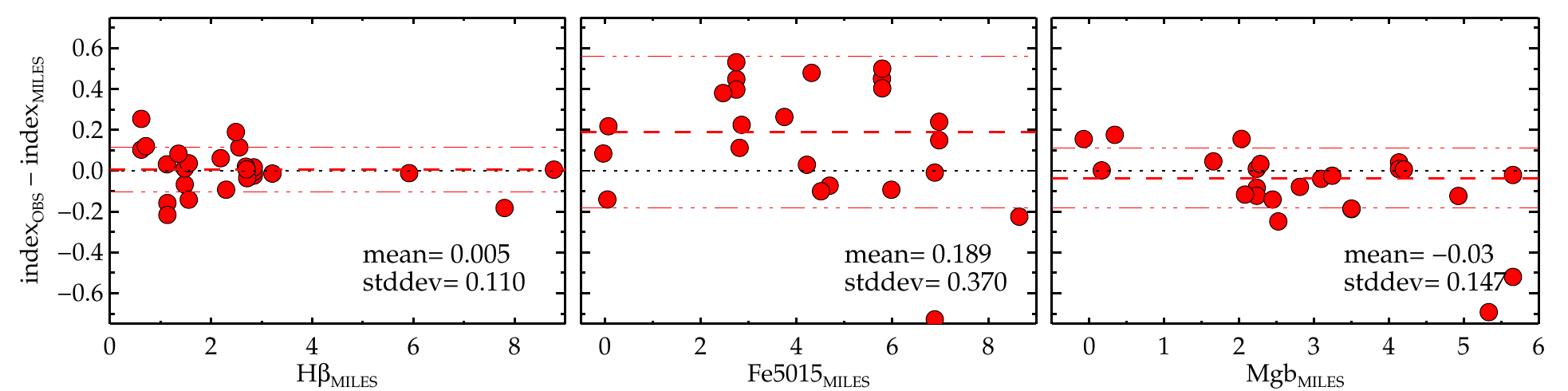}
\caption{The comparison between the line-strength values obtained for the MILES stars we observed and the values for the same stars from the MILES stellar library of \protect\cite{sanchez-blazquez:2006}. The thick dashed red lines show the mean value of the difference between the library and our values, and the thin dotted-dashed red lines indicate the standard deviation. Zero is shown with the dotted black line. All values are in LIS-5.0\,\AA~system of \protect\cite{vazdekis:2010}. See the main text for more details.}
\label{ls-comparison}      
\end{figure*}

When the stellar velocity dispersion becomes smaller than the instrumental resolution, $\sigma \lesssim$100 km/s in the case of SAURON, a high S/N is required to accurately recover $\sigma$ values. This was extensively discussed in, e.g. \cite{toloba:2011} who carried out Monte-Carlo simulations to determine the minimum S/N needed to recover reliable $V$ and $\sigma$. Mimicking the instrumental setup of their data, they created a set of model galaxies for a range of ages and metallicities and found that they were able to recover values with a $\lesssim$10\% error for all their  input $\sigma$ values if the S/N was higher than or equal to~20. \looseness-1

We repeated the exercise of \cite{toloba:2011}, and created a set of simulations to characterize the uncertainties and biases of our velocity dispersion measurements in the low-$\sigma$ and low-S/N regimes. A MILES SSP model \citep{vazdekis:2010} of 5\,Gyr and Solar metallicity (i.e. typical of the average population found in our sample) was used as input galaxy. The spectrum was transformed to mimic the SAURON instrumental setup of our observations, and later convolved to different velocity dispersions and signal-to-noise ratios. The results of this experiment are shown in Fig.~\ref{sigma-sims}. In the left panel we show the potential biases in the recovery of the velocity dispersion for the lowest S/N considered in this work. As expected, the average difference between the input and output values tends to zero as $\sigma_{\rm input}$ increases.  For $\sigma_{\rm input}$ below our pixel scale ($\sim$60 km/s) one can start to appreciate systematic behaviors (i.e. over- and under-estimation of the input values). This effect is, however, well below the uncertainties at the lowest velocity dispersions we measure in our maps  ($\sim$30 km/s). The right panel shows typical uncertainties for a large range in S/N ratios in the velocity dispersion range where the line-of-sight velocity distributions are undersampled. As in the previous plot, at the minimum signal-to-noise level considered for the Voronoi binning any systematic trend is within our measured uncertainties. These tests suggest that our measurements should not be significantly affected by sampling effects.

To directly establish the level of confidence of our results we compared our $\sigma$ measurements with literature values. Figure~\ref{profile_comparison} shows the comparison between our and previously published profiles of \cite{simien:2002}, \cite{geha:2003}, \cite{vanzee:2004}, \cite{beasley:2009}, \cite{chilingarian:2009}, and \cite{toloba:2011}, typically based on high spectral resolution measurements. We plot both the individual bin values, as well as values calculated by averaging the bin values over ellipses (using the kinemetry software of \citealt{krajnovic:2006}). The errors shown are representative of average \textit{bin} errors at a given radius. For most of our objects the agreements is good and the profiles agree. For VCC\,1036 and VCC\,2048 the values generally agree at larger radii but we do not see a drop in $\sigma$ reported by some authors. We do, however, agree to within the errors with the published profiles of VCC\,1261, for which a $\sigma$ drop has also been reported. Certain disagreement between our and published data for VCC\,0929 is very likely caused by the lower quality of the published data, evidenced by large scatter and error bars of those measurements. The overall conclusion from the comparison is that we are able to measure $\sigma$ values which agree to within the errors with most of the profiles available in the literature. Thus, the claim that an accurate determination of $\sigma$ as low as~$\sim$half of the instrumental resolution is possible with high-quality data is now supported not only from a theoretical but also an observational point of view.

\section{Line strengths}
\subsection{Line strengths measurements}

The measurement of absorption line strengths in combination with stellar population models has been used for years to probe the luminosity-weighted age, metallicity and abundance ratios of certain elements in integrated stellar populations. 

For many years now the standard in the field has been the set of 25 indices in the Lick/IDS system \citep{worthey:1994}. Unfortunately, the system has been defined on a low-resolution library (8-11\,\AA), which with the advent of high-resolution spectroscopic data meant that in order to use the indices one had to degrade the resolution of their data to match that of the Lick/IDS system, thus effectively forsaking the data quality. Therefore, a need emerged to define a new system that would avoid this data degradation, as well as other problems inherent to the Lick library, such as changing resolution as a function of wavelength, uncertain calibration of the continuum shape, and low effective S/N due to significant flat-field noise (see e.g. \citealt{worthey:1994} or \citealt{trager:1998}).

The new Line Index System \citep{vazdekis:2010} meets these requirements. It provides constant resolution over the whole wavelength range, and it avoids the above mentioned intrinsic uncertainties associated with the Lick/IDS system. The system is defined at three different resolutions, 5.0\,\AA~(suitable for globular clusters' index comparisons),  8.4\,\AA~(appropriate for studies of dwarf and medium-mass galaxies, roughly matches the Lick/IDS system resolution), and 14.0\,\AA~(designed for studies of massive galaxies). The MILES models are based  on the MILES stellar library presented in \cite{sanchez-blazquez:2006} that contains 985 stars covering a much wider parameter space (T$_{eff}$, log (g), [Fe/H]) -- particularly at lower metallicities --  than any other library available before. Their spectral range is 3540-7610\,\AA~at 2.50~\AA~\citep{falcon:2011a}.\looseness-2

What indices can be measured is mainly determined by the available wavelength range of one's instrument and a given object's redshift.
When observing Virgo objects with SAURON we have 3 traditional Lick indices available: H$\beta$, Mg$b$ and Fe5015, of which H$\beta$ is the  age-sensitive index and the other two are metallicity indicators. In this paper we present line-strength maps and profiles in the LIS-5.0\,\AA~system.

The errors on the indices were calculated through Monte-Carlo simulations. For each realization, we perturbed the spectra to incorporate the following error sources: $V$ measurement, which affects the position of the lines; $\sigma$ measurement, which affects the lines' depth; and systematic errors (in the same way as described earlier for kinematic measurements, using as the error spectrum the difference between the data and the best-fit model spectra).

\subsection{Comparison with published values}

Central integrated line-strength values have previously been published for a number of dEs by \cite{geha:2003}, \cite{michielsen:2008}, and \cite{paudel:2010a}. While we do have a few galaxies in common with their samples, a comparison between their and our line-strength values is not straightforward. The published values are in the old Lick/IDS system and the conversion from Lick/IDS to LIS is rather uncertain and may introduce biases. Moreover, to properly convert we would need to know the exact instrumental setups used to obtain the published values.

\begin{figure*}
\centering
\includegraphics[width=1.00\textwidth]{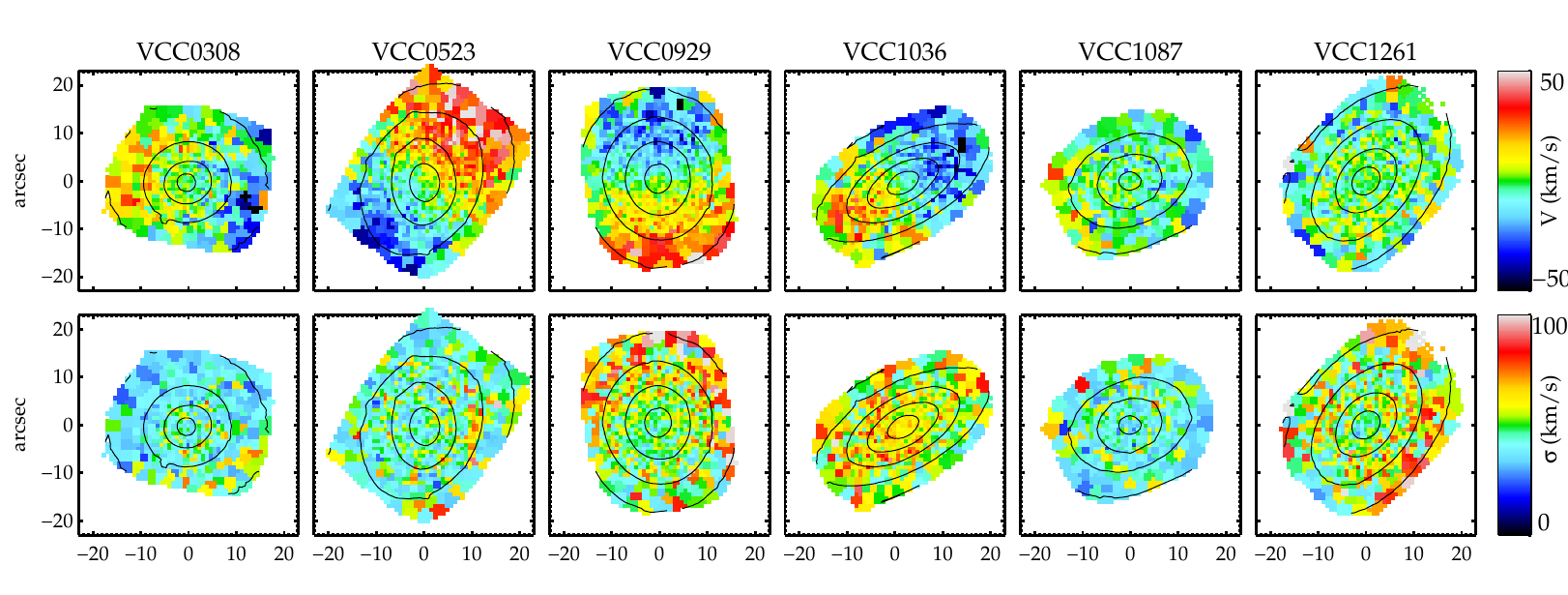} 
\includegraphics[width=1.00\textwidth]{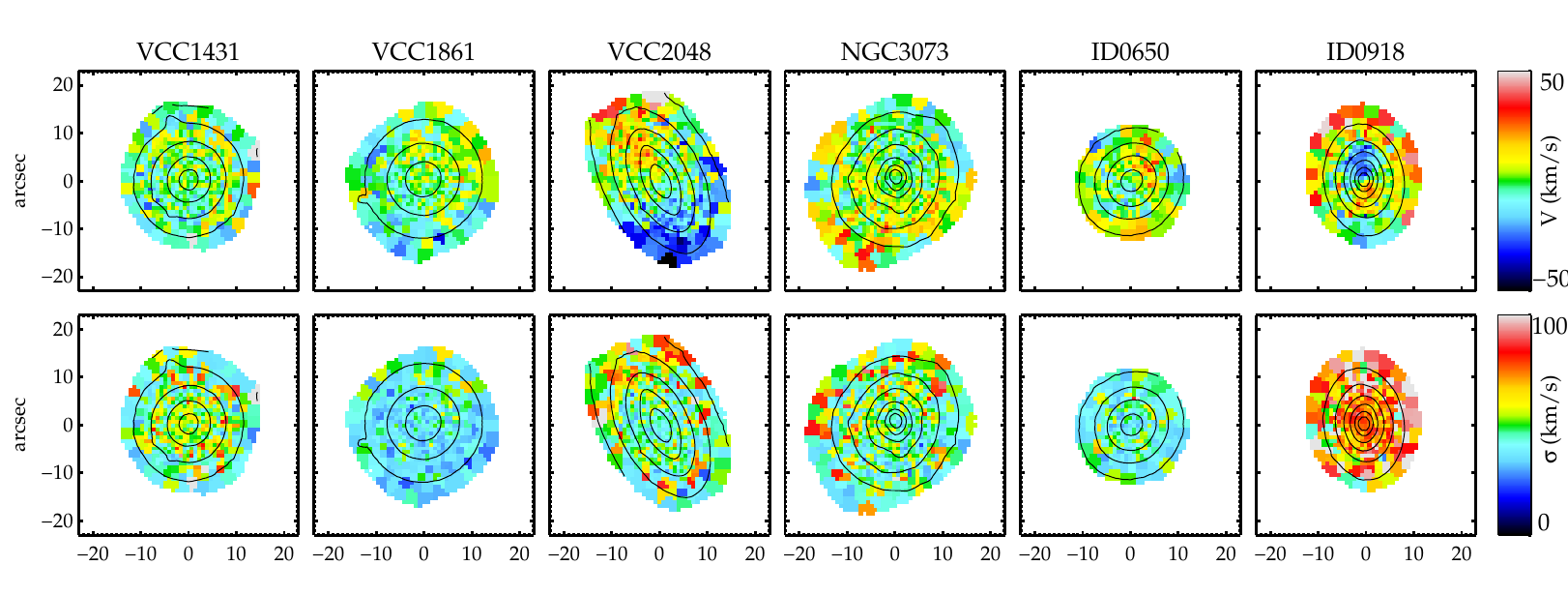}
\caption{Stellar velocity and velocity dispersion maps of our full sample. The maps are overplotted with the isophotes of the reconstructed images and presented on a common scale to facilitate the comparison among galaxies. The same maps (including errors) are presented in Appendix~\ref{app-a} with optimized scales to emphasize the individual characteristics.}
\label{all-kinematics}      
\end{figure*}

We thus opted for a different method to test how reliable our line-strength measurements are. We measured the indices on the MILES stars we observed every night and then compared their values with those of the same stars in the MILES stellar library of \cite{sanchez-blazquez:2006}, measured at our instrumental resolution using the MILES website (\textit{miles.iac.es}). By choosing to test our measurements in this way, we made sure that we had control over all factors to which line-strength measurements are sensitive: the overall data quality, reduction procedures, spectral resolution, and, most importantly, the system in  which line strengths were measured, without the need to resort to conversion tables.

The results of the comparison are in Figure~\ref{ls-comparison} where we show the absolute differences between the observed and the library stars values. The agreement for  H$\beta$ and Mg$b$ measurements is very good, with the average offsets being only 0.005$\pm$0.110\,\AA~and -0.03$\pm$0.147\,\AA, respectively. The offset for Fe5015, though a littler higher (0.189$\pm$0.370\,\AA) is still, like in the case the other two indices, within the error bars of our galaxy observations (see the appendices). The means and standard deviations of the relative errors ($\mathrm{(index_{OBS}-index_{MILES})/index_{MILES}}$, not shown in the figure) are 0.003$\pm$0.056, 0.063$\pm$0.108, and -0.020$\pm$0.054 for H$\beta$, Fe5015, and Mg$b$, respectively. We thus conclude that our measurements are robust.

\section{Results}

With integral-field data we are able to build not just kinematic ($V$ and $\sigma$) and line-strength \textit{profiles}, but \textit{maps} that provide information for every spatially resolved element. This is not the first study devoted to dEs in which integral-field data has been used (see \citealt{prugniel:2005}; \citealt{chilingarian:2009}), however, our maps cover a  $much$ larger area than those few published previously so we can characterize the behavior of kinematic and stellar population parameters out to much (up to four times) larger radii. The available long-slit data do sometimes go a bit farther out (\citealt{chilingarian:2009}, \citealt{toloba:2011}) but the data never reach beyond 25--30''. The spatial information obtained from our data is, thus, much more extensive. Also, when analyzing radial properties, we can average our data points over annuli, which significantly increases S/N and hence reduces error bars. 

Appendix~\ref{app-a} contains a full set of kinematic and line-strength maps, including error maps, for each galaxy. Additionally, in Appendix~\ref{app-b} we plot and tabulate one-dimensional profiles of all measured quantities and their errors in order to enable an easier comparison of our and literature results. Here we show all maps together to emphasize the overall variety and to facilitate the comparison of properties among the galaxies. The rest of this section provides a detailed description of these properties.

\subsection{Stellar kinematic maps}
Figure~\ref{all-kinematics} shows our $V$ and $\sigma$ maps. The scale is the same for all galaxies to emphasize their diversity. The maps show varying degrees of rotation: from $V_{max}\approx$40 km/s for VCC\,0523 to virtually no rotation for VCC\,1261. These rotation values seem to be uncorrelated with flattening since we see all possible combinations of both: flattened rotators, flattened non-rotators, fairly round rotators and, similarly shaped non-rotators. In addition to that, in two galaxies (ID\,0650 and ID\,0918) we observe large-scale kinematically-decoupled components, counter-rotating with regard to the main bodies. The flattened slow rotator VCC\,1087 and non-rotator VCC\,1261 are known (through globular clusters analysis of \citealt{beasley:2009}) to display strong rotation at much larger radii (4--7 $R_e$).  The rotation in the (almost) round VCC\,0308 and VCC\,0929 could be due to them being seen not perfectly face-on, or it could be ascribed to their specific orbital configuration if they were triaxial. We see kinematic twists in two systems: in VCC\,0523, known to host a bar \citep{lisker:2007} and showing weak signs of spiral arms \citep{lisker:2009}, it is very clear that the central photometric isophotes are twisted with regard to the outer ones, and it is the outer ones that agree with the kinematic major axis of the galaxy; in ID\,0918 the twist is less pronounced, but has similar characteristics.

The velocity dispersion maps are rather flat, in the $\sigma \simeq$40-60 km/s range, with shallow gradients for a few of the galaxies: a slight central increase (VCC\,1431) or decrease (VCC\,0929, VCC\,1261). Only for 1 field galaxy is $\sigma$ significantly higher (ID\,0918, $\sim$80 km/s). The interpretation of flat dispersion profiles requires employing dynamical models and will be presented in our Paper II.

What is most significant in these stellar kinematic maps is that no two galaxies in our sample are really alike. The implication of this to their formation histories is discussed in Section~\ref{section:discussion}.

\begin{figure*}
\centering
\includegraphics[width=1.00\textwidth]{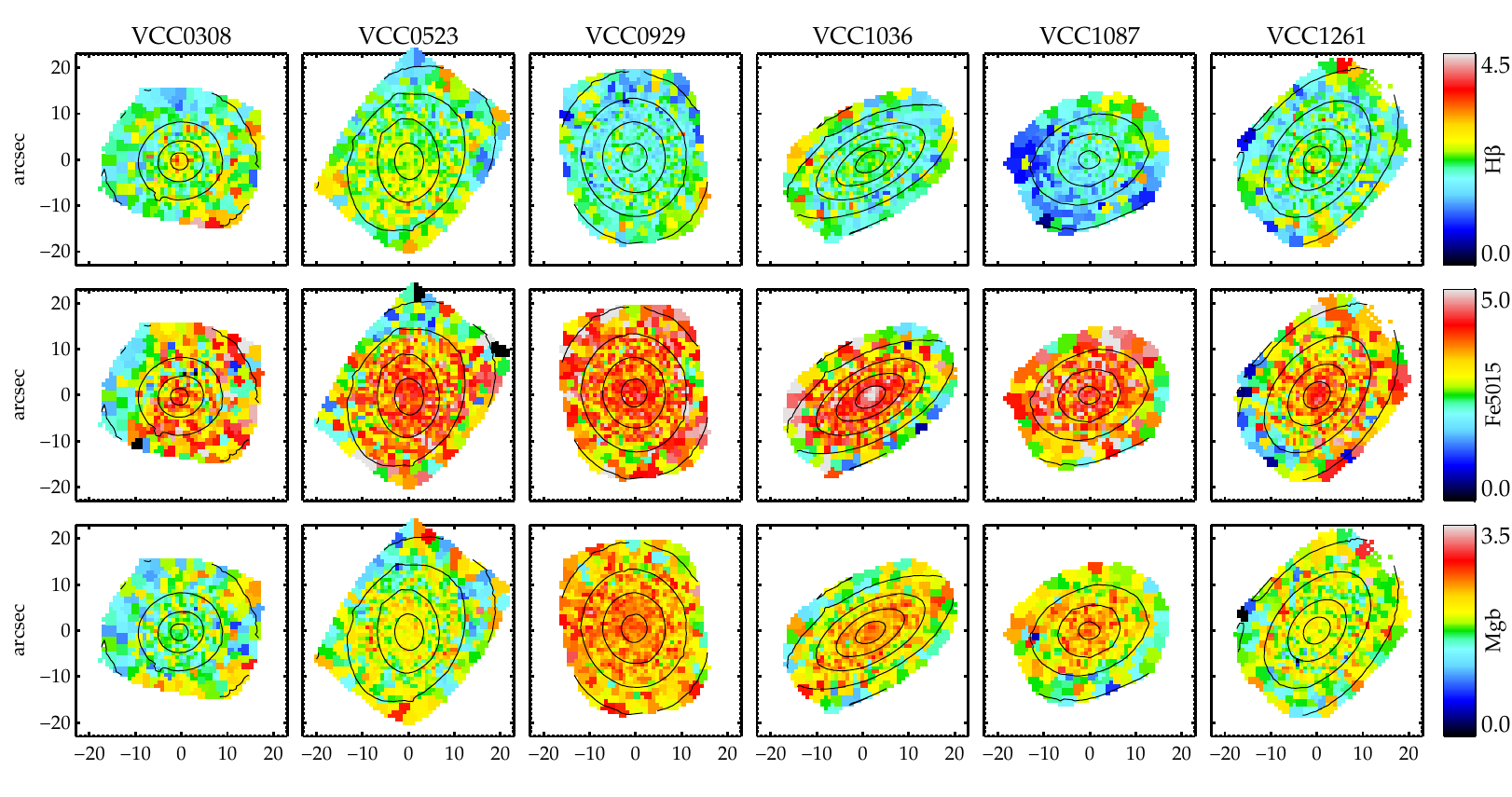} 
\includegraphics[width=1.00\textwidth]{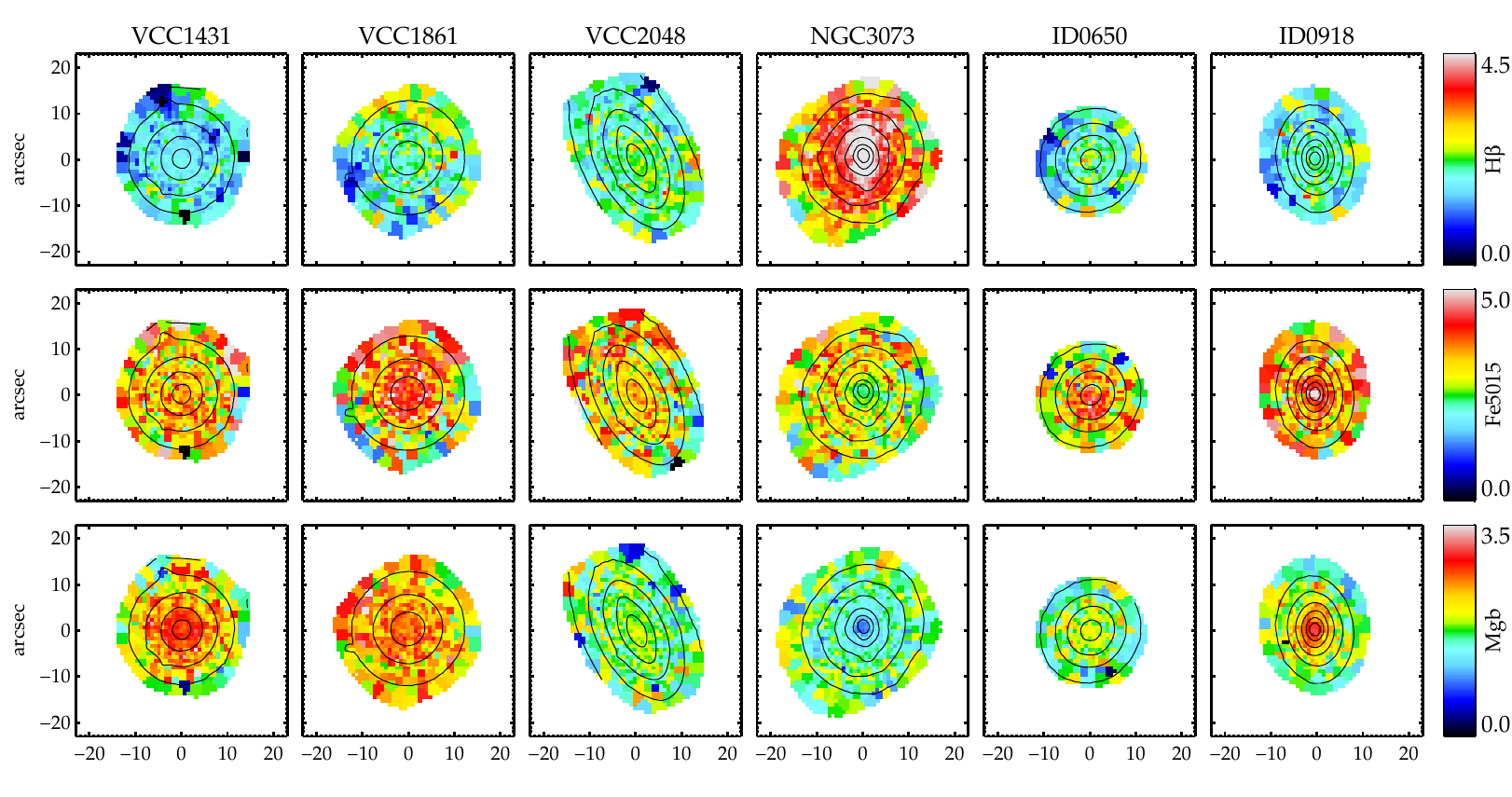}
\caption{Line-strength maps of our full sample. The maps are overplotted with the isophotes of the reconstructed images and presented on a common scale to facilitate the comparison among galaxies. See Appendix~\ref{app-a} for maps with individually optimized scales and errors.}
\label{all-lines}      
\end{figure*}

\subsection{Line-strength maps}
The goal of the stellar population analysis is to gain insight into the star-formation and assembly history. For example, two-dimensional data makes it possible to connect stellar populations with kinematic structure, as shown by \cite{kuntschner:2006} where various discrepancies between isophotal and isoindex contours were found for their sample of giant elliptical (E) and lenticular (S0) galaxies. 

We do not seem to find such strong signatures in our data (Figure~\ref{all-lines}), i.e. our data are consistent with no deviation between isophotes and isoindex contours with the  exception of one galaxy (VCC\,1036, Fe5015 map). This would suggest that the fast-rotating components, i.e. disks, are not different from the main bodies in terms of populations. However, the gradients in the indices do differ among our galaxies: for Fe5015 and Mg$b$ (i.e. the metallicity tracers), while negative (i.e. more metal-rich in the center) in all galaxies but NGC\,3073 , they do vary from almost flat to having a strong peak (or in the case of NGC\,3073: a dip) in the center; H$\beta$ (age tracer) gradients are similar in that respect, some showing a highly-concentrated peak (like in the blue-core VCC\,0308), with others being essentially flat. The general trend we get from our maps is that dEs are on average older and more metal-poor in their outer parts, in agreement with \cite{michielsen:2008}. We further investigate these points in Paper III, where a detailed stellar population analysis, including population gradients and star formation histories (using full spectrum fitting techniques) will be presented, and in Paper IV,  where we use aperture measurements to analyze global scaling relations and compare dEs to other galaxy types.

\subsection{Notes on individual galaxies}

Here we provide a brief description of the properties of our maps for each galaxy. The following abbreviations will be used throughout the section: 1)~\cite{pedraz:2002}: P02, 2)~\cite{simien:2002}: S02, 3)~\cite{geha:2003}: G03, 4)~\cite{vanzee:2004}: Z04, 5)~\cite{lisker:2006a}: L06, 6)~\cite{chilingarian:2009}: C09, and 7)~\cite{toloba:2011}: T11. 

\textit{VCC\,0308} is designated dE(bc,di) in the notation of L06, and it is the only blue-core (bc) galaxy in our sample. The H$\beta$ map shows a strong peak in the center, suggestive of the presence of a younger population, consistent with the '(bc)' classification. It is a round galaxy in the outskirts of the cluster, showing a mild level of rotation, signs of which have also been seen by T11. 

\textit{VCC\,0523} is a mildly flattened, nucleated dwarf in the inner parts of the cluster; L06 found a disk and a bar in this galaxy based on photometric decomposition, so the galaxy is denoted dE(di). We confirm this result by showing how the photometric and the kinematic major axes of the galaxy are misaligned due to the bar. The galaxy is rotationally supported, in agreement with T11 results from their mean velocity and velocity dispersion curves. 

\textit{VCC\,0929} is a fairly round galaxy with a high level of rotation. Previous results also indicate rotation, though to a lesser degree (P02) or showing large uncertainties (S02). Line-strength maps are essentially flat, showing only a mild increase towards the center in the Fe5015 and Mg$b$ maps.

\color{black}{}

\textit{VCC\,1036}, of the dE(di) type, is a very flat object that lies in the central parts of the cluster. It is a well-studied galaxy, previously analyzed by S02, G03, Z04, and C09. In agreement with these studies we find that the galaxy is fast-rotating. We do not confirm the presence of a central $\sigma$~drop reported by C09. The H$\beta$ map is flat, and Fe5015 map shows a central increase/peak, which is yet more pronounced in the Mg$b$ map.

\textit{VCC\,1087}, centrally located, is another well-known flattened nucleated dwarf, this time, however, with no substructures (L06) and little apparent rotation as reported previously by S02, G03, C09 and T11. Our maps do show some level of rotation at the edge of the field, it is, however, almost consistent with zero. Interestingly, \cite{beasley:2006} results from the analysis of globular clusters located at up to 4--7 effective radii reveal a significant level of rotation at those larger distances. It is, thus, possible that our maps look flat due to insufficient radial coverage. 

\textit{VCC\,1261} is the only galaxy in our sample, which, despite significant flattening, truly exhibits no signs of rotation out to 1 effective radius (previously noted by G03, Z04, C09 and T11). C09 velocity curve and sigma profile suggest a possible presence of a small kinematically decoupled component. Our data, as well as those of G03 and T11, do show some signs of a central $\sigma$~drop, though it is more spatially extended and not as deep. Our line-strength maps show a central peak in the H$\beta$ and Mg$b$ maps, suggestive of a rather strong positive age and negative Z gradient. The galaxy's GCs were studied by \cite{beasley:2009} with the similar findings as for VCC\,1087 above. \looseness-1

\textit{VCC\,1431} is a very round, non-rotating galaxy in the center of the cluster. The galaxy has previously been analyzed by T11. We note a small increase in $\sigma$ towards the galaxy's center, which is however, consistent with zero to within the errors. The line-strength maps show a clear central peak in the Fe5015 map, while the H$\beta$ and Mg$b$ maps are flat.

\textit{VCC\, 1861} is an almost perfectly spherical non-rotator (with an average $V<$10 km/s), but it must be remembered that, like in the case of VCC\,1431 above, we may be viewing the galaxy face-on, in which case we would be unable to detect any (possible) rotation. No substructures are present as reported by L06. Our line-strength maps show a small negative Fe5015 and Mg$b$ gradient.

\textit{VCC\,2048} is a flattened disky rotator situated in the cluster's outskirts. Previous kinematic analyses are found in S02 and C09. The latter finds a pronounced $\sigma$~drop in the center. The drop would be undetectable for us at the claimed minimum $\sigma$ level (20~km/s).  Our line-strength maps seem fairly flat, with the exception of the Mg$b$ map which shows a slight increase towards the center. VCC\,2048 is the galaxy considered by Kormendy \& Bender (2011) to be a transition object and a ``missing link'' between the S0 and dE (Sph in their notation) classes.

\textit{NGC\,3073} is a field galaxy in the vicinity of the larger NGC\,3079 (with their projected separation of $\sim$10 arcmin). The galaxy has previously been observed by  T11 and \cite{emsellem:2011}. The classification of this galaxy varies from source to source: SAB0$^-$ in \cite{vaucouleurs:1991}, S0 in \cite{nilson:1973}, and T11 classify it as dE(di). The galaxy's integrated intensity isophotes are distorted by the presence of gas which is later seen in the spectra in the form of strong emission lines . \cite{serra:2011} show the H{\sc i} pattern for NGC\,3073 where a tail is present pointing away from its large neighbor. Small level of rotation was previously detected by S02 and T11, and is indeed confirmed here.

\textit{ID\,0650} is a truly isolated field galaxy, with no galactic neighbors within $\sim$30 arcmin projected radius. This is the first kinematic data for this object available in the literature. We find a large-scale counter-rotating component with an approximate diameter of 12\,arcsec (or $\approx$1.5 kpc). The Fe5015 and Mg$b$ maps show a slight central peak. 

\textit{ID\,0918} is a field galaxy with the closest galactic companion at a projected distance of $\sim$20 arcmin, also with no kinematic data available so far. The integrated intensity isophotes show a slight misalignment of the inner and outer parts, the latter agree with the galaxy's rotation pattern. We report a discovery of a strong counter-rotating component, whose approximate size is 14\,arcsec (or $\approx$1.1~kpc) along the major axis. The line-strength maps show a strong central peak in the Fe5015 and Mg$b$ maps.

\section{Discussion and conclusions}\label{section:discussion}

\begin{figure}
\centering
\includegraphics[width=1.00\columnwidth]{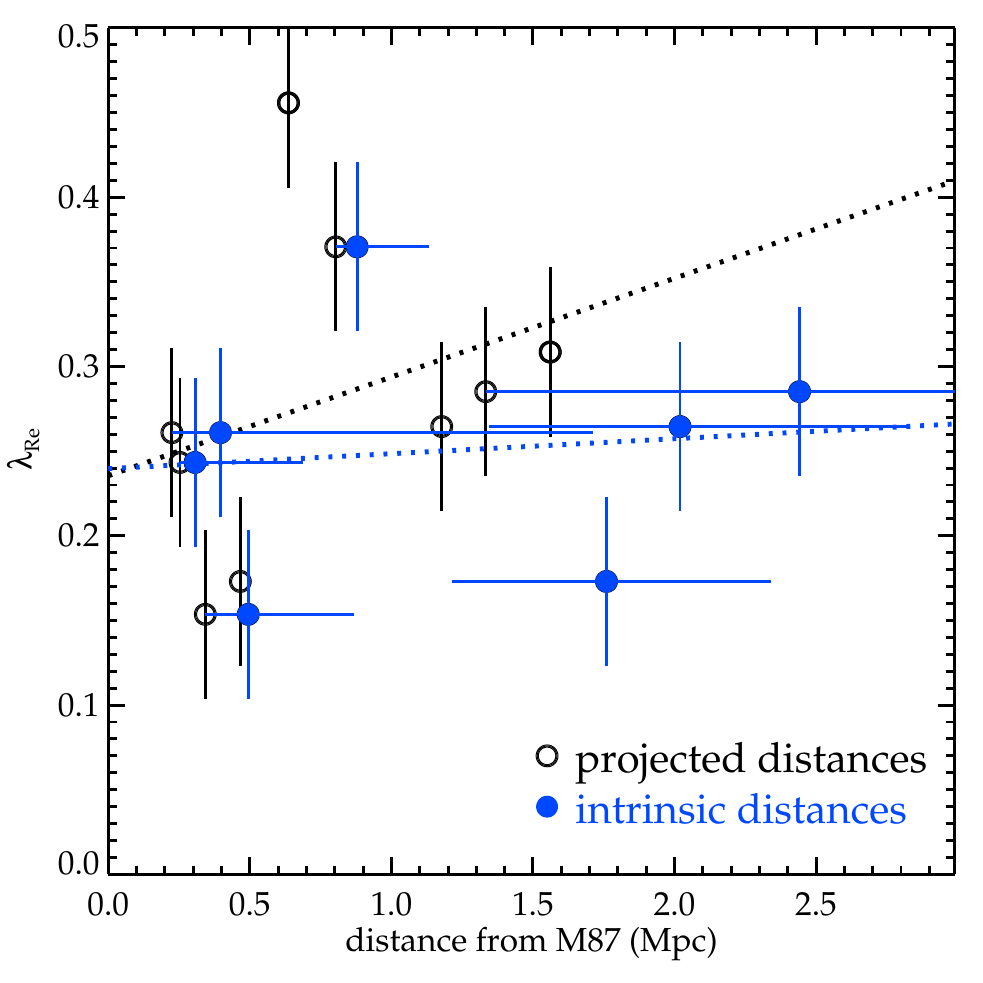} 
\caption{Integrated (within 1 effective radius) specific angular momentum $\lambda_R$ versus projected (open black circles) and deprojected (filled blue circles) Virgocentric distance. The errors on $\lambda_{Re}$ are assumed to be 0.05 after \protect\cite{emsellem:2011} (see their Section 4.5). The dotted lines show the fits to the data for both types of distances. While a weak trend is seen when the projected distances are used, this is not true for the intrinsic deprojected distance. No relation between the rotational support and location in the cluster is thus seen in our data.}
\label{lambdar}      
\end{figure}

We have presented stellar kinematic and absorption line-strength maps for a sample of 12 Virgo Cluster and field dwarf elliptical galaxies. This is to date the largest sample of dEs for which integral-field data is available, and offers the largest spatial coverage so far. We have shown how having high S/N data and performing careful data reduction enables one to recover velocity dispersion values well below instrumental resolution. We do not discuss here all the relations presented in the Introduction since this will be done in more detailed subsequent papers that will each focus on different aspects of dEs. Instead, we outline some of the most noteworthy aspects of the data and focus on the meaning of the objects' structural variety.

Even having a relatively small sample we still find all possible combinations of flattening and rotation in our kinematic maps, together with twists and decoupled components. The maps not only show an overall remarkable variety of properties but also draw our attention to interesting individual cases. Our flattened slow/non-rotators (VCC\,1087 and VCC\,1261) deserve particular attention since the analysis of their globular clusters systems (\citealt{beasley:2006} and \citealt{beasley:2009}) shows rapid rotation at larger radii and, in one case, a slight misalignment of axes between the GC and stellar components. One possible explanation could be that the initially present angular momentum has migrated outwards, e.g. due to a disk--halo coupling, still, the actual processes are difficult to name. 

The kinematically-decoupled components (KDCs) found in ID\,0650 and ID\,0918 are similar in physical size to those found in giant elliptical galaxies of \cite{krajnovic:2011}. KDCs in dwarfs are more difficult to explain given that we generally believe they have not formed through mergers \citep{derijcke:2004}. For cluster members there exist studies that suggest the possibility of creating a KDC via harassment \citep{gonzalez-garcia:2005}. Still, for isolated galaxies like ID\,0650 and ID\,0918 the puzzle remains unsolved, since ram-pressure stripping, though believed to be able to act even on objects outside of dense environments (especially if concepts like warm-hot intergalactic medium are invoked, e.g. \citealt{dave:2001}, which would allow for more efficient stripping), is nevertheless unable to sufficiently alter the stellar composition. It is yet unclear whether the peaks in the Fe5015 and Mg$b$ maps are related to the presence of the KDCs, i.e. whether stellar-populations-wise the KDCs could be distinct from the main bodies of the galaxies (e.g. external gas accretion can be responsible for creating counter-rotating components). All these findings invite further investigation, in particular of the orbital make-up of these systems. 

The line-strength maps also show a significant degree of diversity. The gradients visible in the maps vary from flat to strongly peaked in the centers, showing a spread in stellar ages and metallicities, both \textit{among} and \textit{within} the galaxies. All but one systems are essentially gas-free, indicative of no recent or ongoing star formation (the exception is one of the field galaxies, NGC\,3073). Regardless of the variety, and as has already been found by \cite{michielsen:2008}, the comparison of our and \cite{kuntschner:2010} integrated line-strength values shows that the ages of dEs are on average younger, and the metallicities lower than those of giant early-type galaxies (ETGs, \citealt{rys:2012jenam}). What is worth noting is that, unlike for a large number of giant early-type galaxies of \cite{kuntschner:2006}, no deviations are found between isophotes and isoindex contours of our dwarfs. This could suggest that any kinematic structures have \textit{not} been built up by star formation activity restricted to a certain region of a given galaxy. We will explore these questions in more detail in a later paper of this series.

To find out whether the presence of photometric and kinematic (sub)structures is correlated in any way, we have made one-to-one comparisons of  our kinematics with the results of Janz et al. (2012b, in prep.). We find that among multiple component objects we have both rotators and non-rotators, the same goes for single-component galaxies; also, two of our slow-rotators contain lenses. Additionally, four out of five rotators are disky galaxies, based on the \cite{lisker:2007} nomenclature. It thus seems that the presence of photometric and kinematic (sub)structures is not correlated in any obvious way, which yet again strengthens the variety argument.

We note here that this variety combined with the variety found by \cite{krajnovic:2011} in the ATLAS3D ETGs should not be viewed as a sign of dEs and giant ellipticals (Es) being continuous in their properties. Namely, 70\% of the galaxies included in the ATLAS3D sample are S0 galaxies and they have been shown to be fundamentally different from the other ETG type, Es, in both their kinematic (\mbox{ATLAS3D} and SAURON surveys) and photometric (\citealt{kormendy:2012}) properties, with S0s (as defined photometrically) and fast rotators (as defined kinematically)\footnote{The terms are \textit{not} interchangeable, as discussed in \cite{emsellem:2011}.} having properties more similar to those of late-type galaxies. If indeed S0s are faded spirals, as suggested most recently by \cite{kormendy:2012}, then the fact that our dEs show a similar level of variety as the ATLAS3D fast rotators only supports the claim of them being transformed late-type galaxies. 

Our results are, thus, best understood in the context of formation scenarios that, first, assume an infall of younger, late-type objects as the source of dE progenitor galaxies, and second, allow for the random shaping of galaxy properties, based on the variety initial infall parameters (disk inclination and orbit type) and the number of encounters with large galaxies. The combined influence of ram-pressure stripping and harassment fulfills these requirements, still, the exact impact of the two effects is not yet known (see, e.g. the simulations of \citealt{smith:2012} which suggest that the removal of the gas potential could be the key source of disk heating). The tell-tale signs of the environmental influence can be found in changing galaxy properties as a function of clustrocentric distance, as discussed in the Introduction. 

What we would like to stress, however, is that the analysis of any distance trend needs to be performed with caution, in particular, the depth of a cluster needs to be taken into account since using projected (instead of physical) distances might yield biased results, especially when dealing with incomplete samples. For example, we do not find a confirmation for the \cite{toloba:2009} claim that rotationally-supported galaxies are preferentially located in the cluster outskirts. To determine the level of rotational support we used the specific angular momentum $\lambda_{Re}$ of \cite{emsellem:2007}\footnote{While we could reproduce the (v/$\sigma$)* from \cite{toloba:2009}, we choose not to do so, following the warning of \cite{cappellari:2007}, see their Section 6.3.}, where $Re$ signifies values integrated within 1 effective radius, and then plotted it as a function of deprojected clustrocentric distance. The relation we find is essentially flat. The same plot using \textit{projected} distances (not shown here) shows a slight upward trend, i.e. would be in agreement with \cite{toloba:2009}. Still, our sample size prohibits any firm conclusions.

It is also important to note that, while we expect to find all these trends relations of properties with distance, we also \textit{expect} to see  scatter around them, including counter examples. This is because transformation by repeated encounters is, to a certain degree, a random process, therefore there will always be galaxies whose characteristics have been greatly altered in an individual encounter of a specific configuration that, for example, unexpectedly creates a non-rotating galaxy in the outskirts of a cluster, a rotating one close to the center, a fairly round, rotating object, etc.

Our findings fit this picture remarkably well. Unlike other authors (e.g. \citealt{lisker:2008}, \citealt{koleva:2009}, \citealt{paudel:2011}) we believe that the presence of a (large) variety of properties does not necessarily imply different formation channels. Since both ram-pressure stripping and harassment are stochastic processes, a large spread in properties is even \textit{expected}. Thus, we consider the variety of characteristics we find in this work to be our most important result. \looseness-1

The detailed analysis of relations of dEs to other galaxy types, their dynamical structure, and their star-formation histories is an attempt at providing answers to the question of the relative importance of various proposed transformation mechanisms and will be presented in the forthcoming papers. It would, in addition, be very valuable to study in two dimensions the properties of larger samples of irregulars, dwarf spirals, and field dEs, to compare them with cluster dEs.

\section*{Acknowledgments}
\small
We would like to thank Anne-Marie Weijmans for the help with the early stages of the data reduction, Elisa Toloba, Reynier Peletier and Mina Koleva for useful suggestions and comments throughout the development of this project, Thorsten Lisker for the comments on the draft version of this paper, and Joachim Janz for providing a draft of his paper ahead of publication. AR thanks John Kormendy for fruitful discussions during the XXIII Canary Islands Winter School of Astrophysics in November 2011. We thank the anonymous referee for their useful feedback and constructive comments that have helped optimize the presentation of this work. AR acknowledges the repeated hospitality of the Max Planck Institute for Astronomy in Heidelberg, to which collaborative visits contributed to the quality of this work. JFB acknowledges support from the Ram\'{o}n y Cajal Program financed by the Spanish Ministry of Economy and Competitiveness (MINECO). This research has been supported by the Spanish Ministry of Economy and Competitiveness (MINECO) under grants AYA2010-21322-C03-02 and AIB-2010-DE-00227. This research has made use of the NASA/IPAC Extragalactic Database (NED) which is operated by the Jet Propulsion Laboratory, California Institute of Technology, under contract with the National Aeronautics and Space Administration. The paper is based on observations obtained at the William Herschel Telescope, operated by the Isaac Newton Group in the Spanish Observatorio del Roque de los Muchachos of the Instituto de Astrofísica de Canarias.
\normalsize

\bibliography{biblio}

\appendix
\section{Maps}
\label{app-a}
Here we provide maps of intensity, $V$, $\sigma$, and line strengths for the 3 indices, H$\beta$, Fe5015, and Mg$b$, in the LIS--5.0\,\AA~system. Each map is accompanied by an error map, obtained through Monte-Carlo simulations as described in section \ref{kininfo}. The plots are complemented with basic galaxy characteristics, repeated here from Table 1 for the reader's convenience. The plotting ranges are adjusted to emphasize individual features.

\section{Profiles}
\label{app-b}
In order to facilitate the comparison between our and literature data, we plot here (Figure~\ref{all-profiles}) one-dimensional (1D) profiles of all our measured quantities, mimicking long-slit data. Tabulated values together with errors are provided in Table~\ref{1Dprofiles-1}. The profiles were extracted along ellipses in steps of 1'' using the \textit{kinemetry} software of \cite{krajnovic:2006}. The listed errors were obtained by running the software on the error maps presented in Appendix~\ref{app-a}, so they reflect average $bin$ errors at a given elliptical distance.

% \newpage
% \setcounter{figure}{1}
\renewcommand{\thefigure}{\Alph{figure}\arabic{subfigure}}
\addtocounter{subfigure}{1}

\clearpage
\begin{figure*}
\begin{center}
  \includegraphics[width=1.25\textwidth,angle=90]{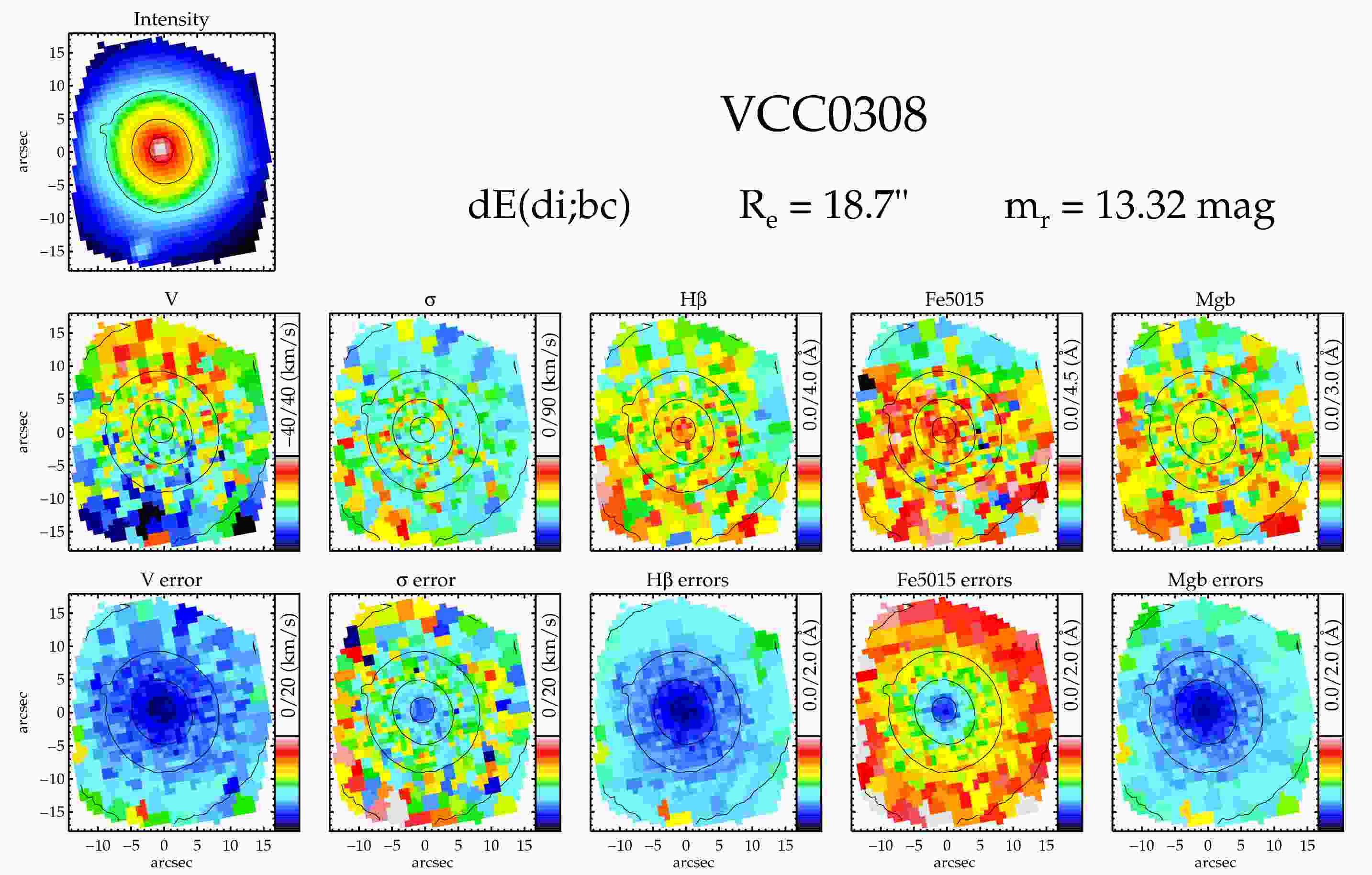}
\end{center}
\caption{Maps of intensity, $V$, $\sigma$, and line strengths for the 3 indices: H$\beta$, Fe5015, and Mg$b$, accompanied by error maps. See Appendix~\ref{app-a} for more details.}
\label{ppxf-maps0308}
\end{figure*}

\addtocounter{figure}{-1}
\addtocounter{subfigure}{0}

\clearpage
\begin{figure*}
\begin{center}
  \includegraphics[width=1.25\textwidth,angle=90]{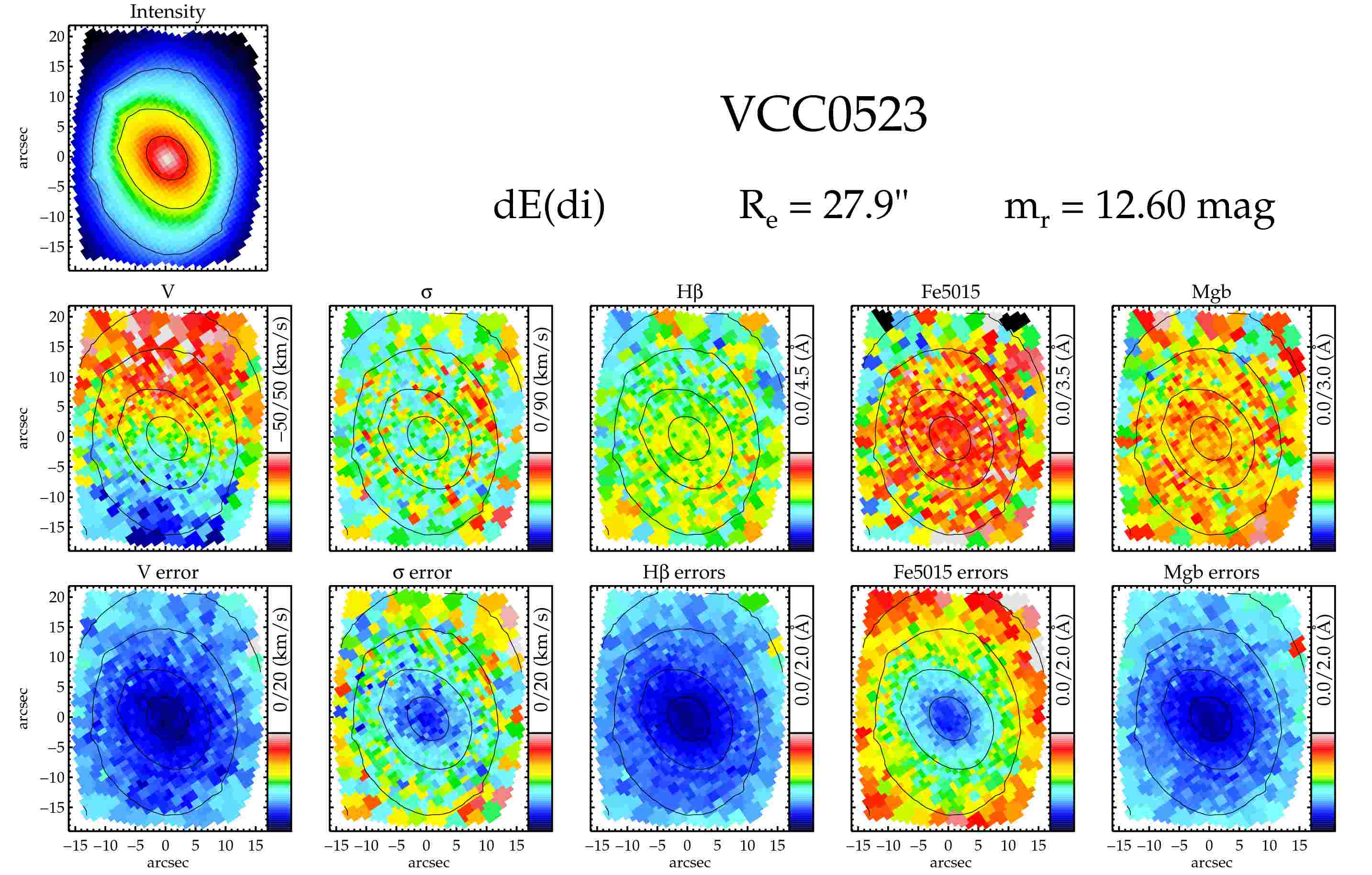}
\end{center}
\caption{Continued.}
\label{ppxf-maps0523}
\end{figure*}

\addtocounter{figure}{-1}
\addtocounter{subfigure}{0}

\clearpage
\begin{figure*}
\begin{center}
  \includegraphics[width=1.25\textwidth,angle=90]{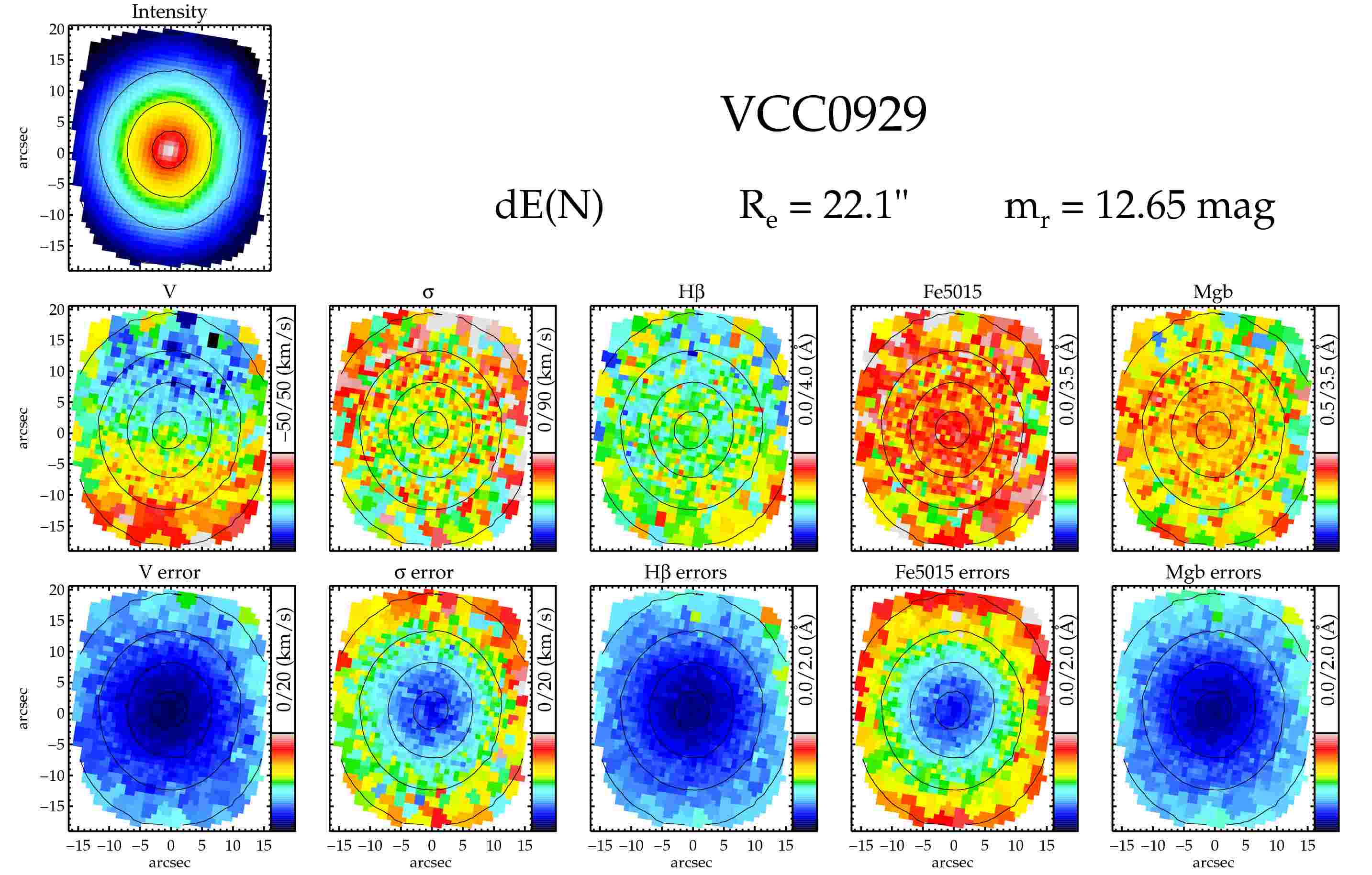}
\end{center}
\caption{Continued.}
\label{ppxf-maps0929}
\end{figure*}

\addtocounter{figure}{-1}
\addtocounter{subfigure}{0}

\clearpage
\begin{figure*}
\begin{center}
  \includegraphics[width=1.25\textwidth,angle=90]{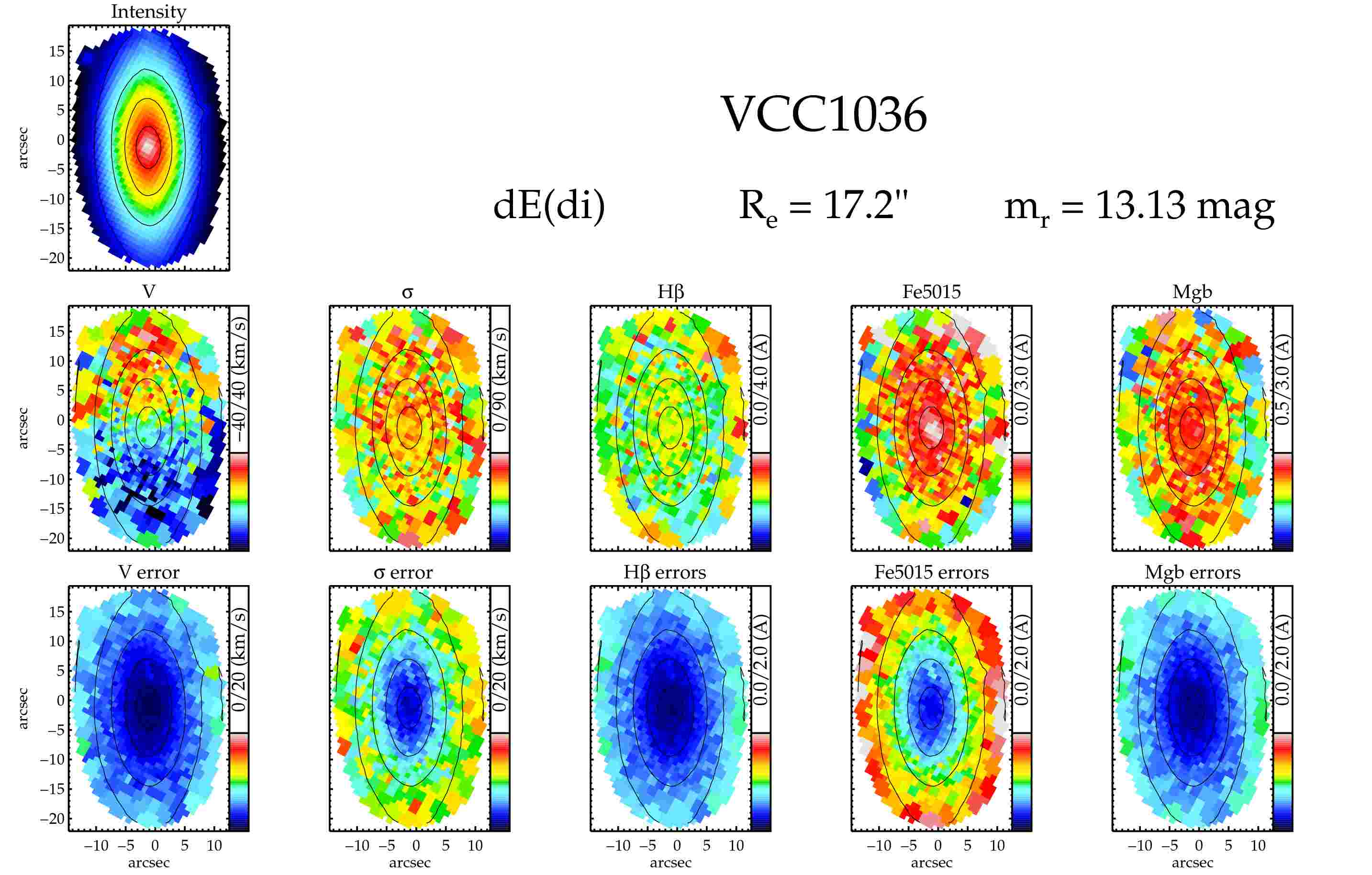}
\end{center}
\caption{Continued.}
\label{ppxf-maps1036}
\end{figure*}

\addtocounter{figure}{-1}
\addtocounter{subfigure}{0}

\clearpage
\begin{figure*}
\begin{center}
  \includegraphics[width=1.25\textwidth,angle=90]{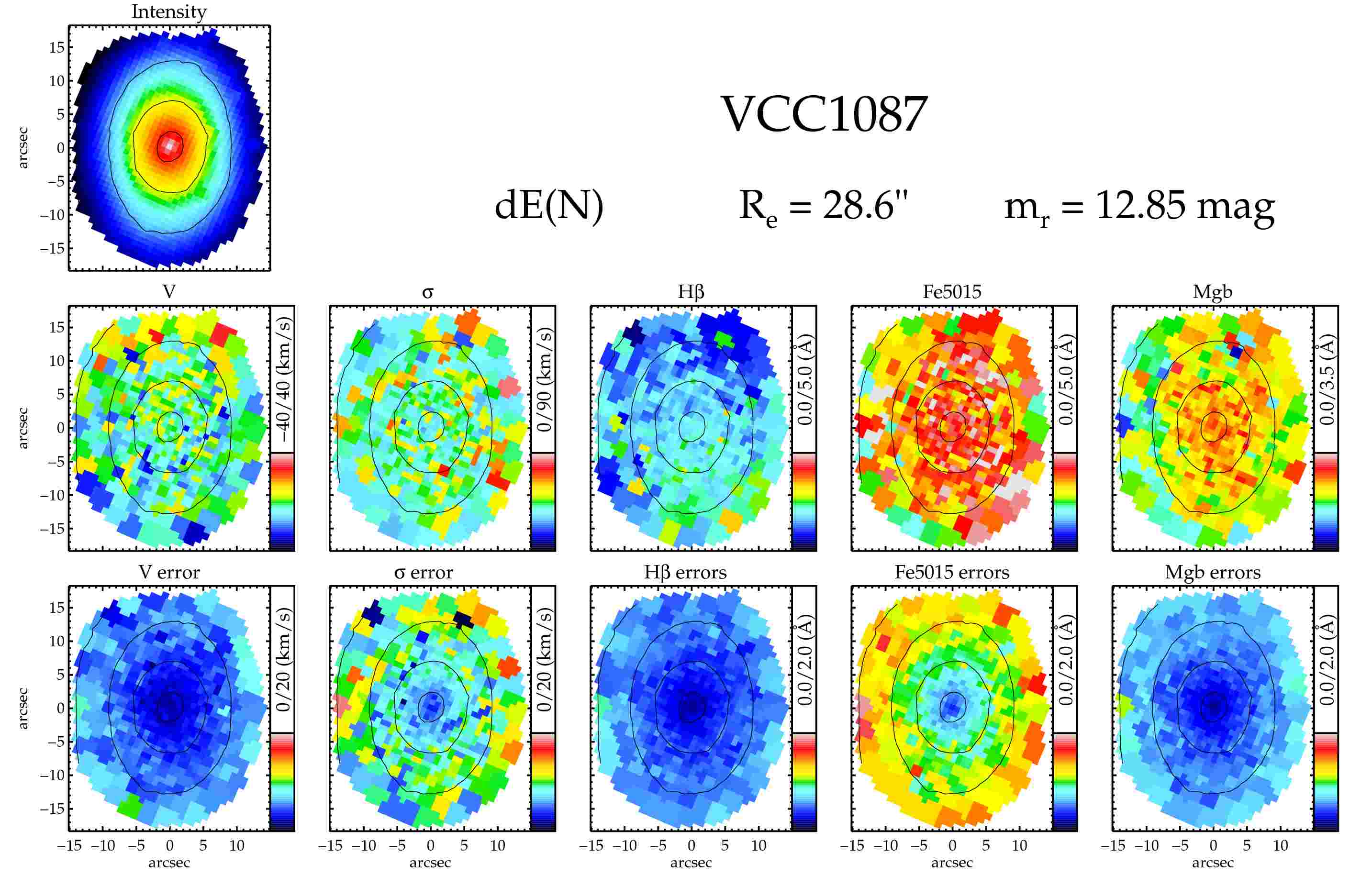}
\end{center}
\caption{Continued.}
\label{ppxf-maps1087}
\end{figure*}

\addtocounter{figure}{-1}
\addtocounter{subfigure}{0}

\clearpage
\begin{figure*}
\begin{center}
  \includegraphics[width=1.25\textwidth,angle=90]{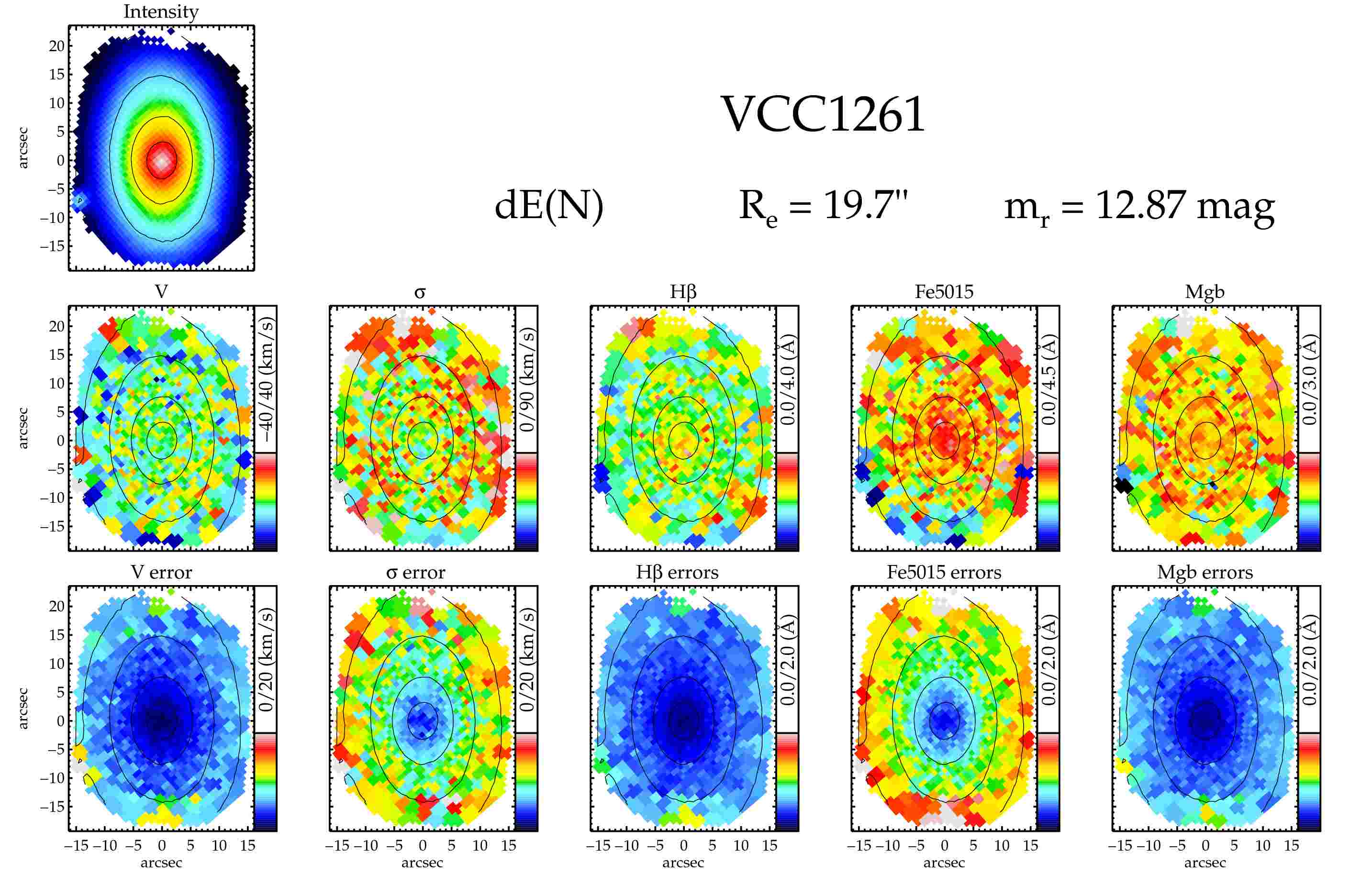}
\end{center}
\caption{Continued.}
\label{ppxf-maps1261}
\end{figure*}

\addtocounter{figure}{-1}
\addtocounter{subfigure}{0}

\clearpage
\begin{figure*}
\begin{center}
  \includegraphics[width=1.25\textwidth,angle=90]{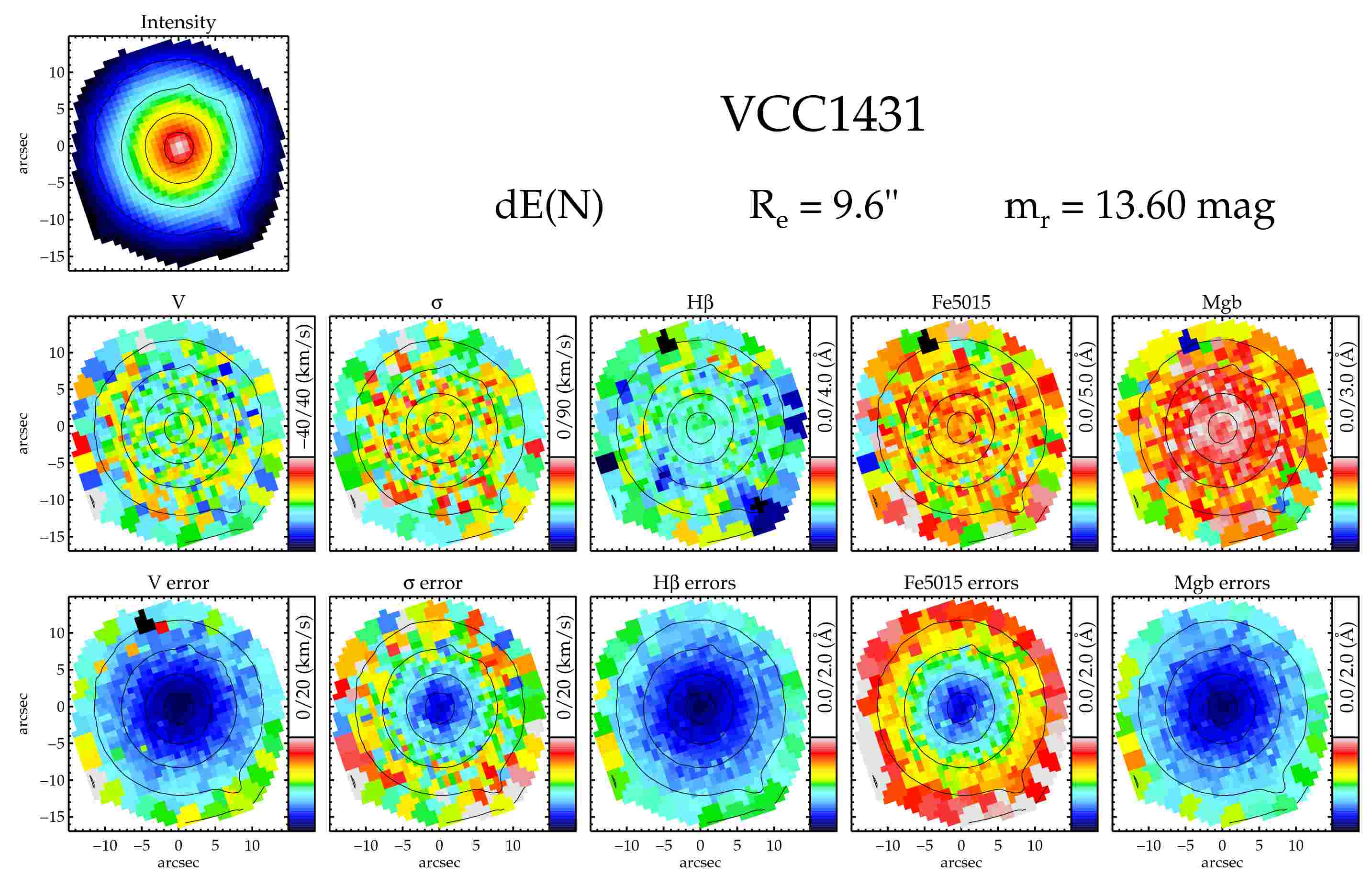}
\end{center}
\caption{Continued.}
\label{ppxf-maps1431}
\end{figure*}

\addtocounter{figure}{-1}
\addtocounter{subfigure}{0}

\clearpage
\begin{figure*}
\begin{center}
  \includegraphics[width=1.25\textwidth,angle=90]{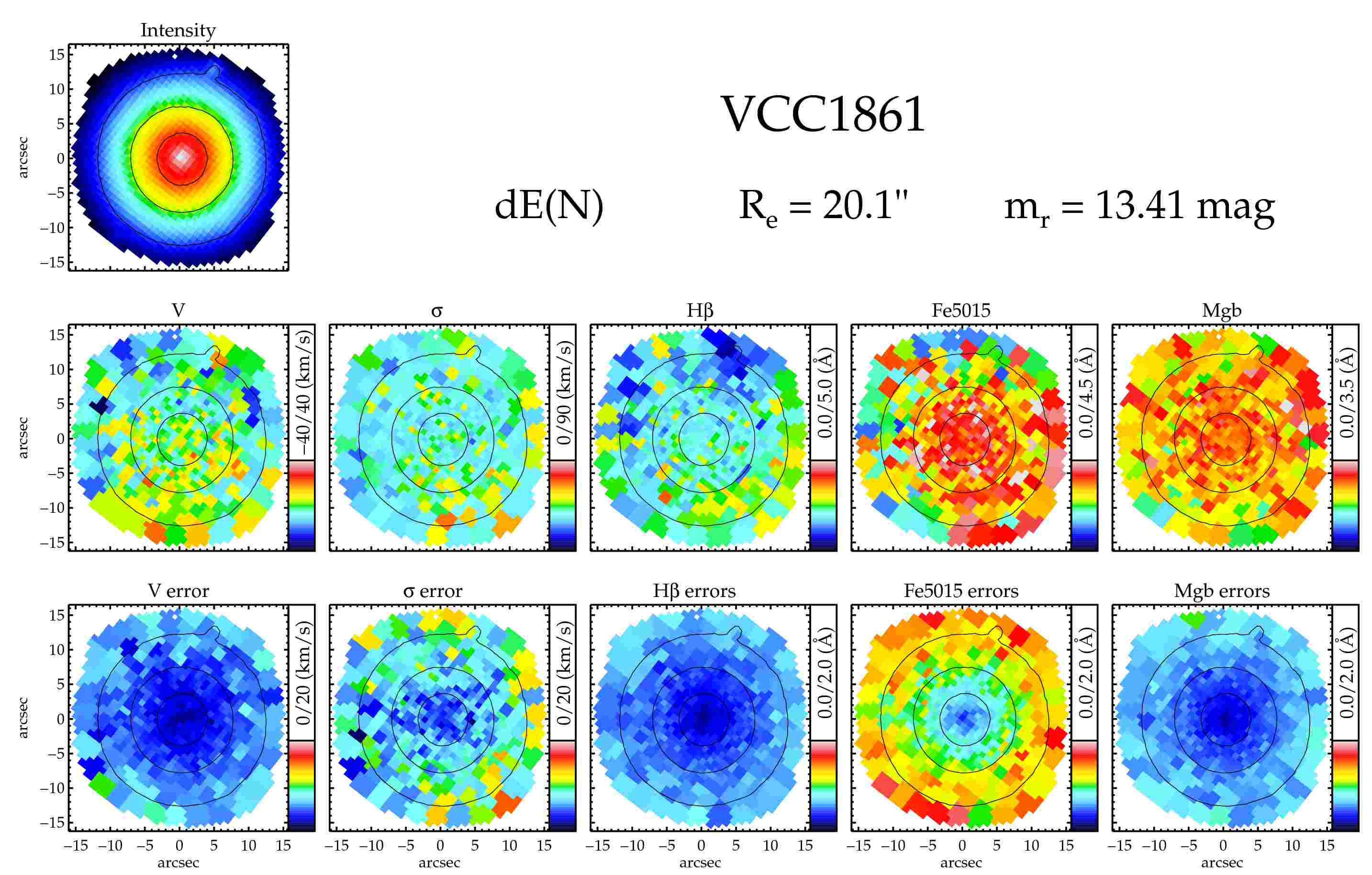}
\end{center}
\caption{Continued.}
\label{ppxf-maps1861}
\end{figure*}

\addtocounter{figure}{-1}
\addtocounter{subfigure}{0}
\clearpage
\begin{figure*}
\begin{center}
  \includegraphics[width=1.25\textwidth,angle=90]{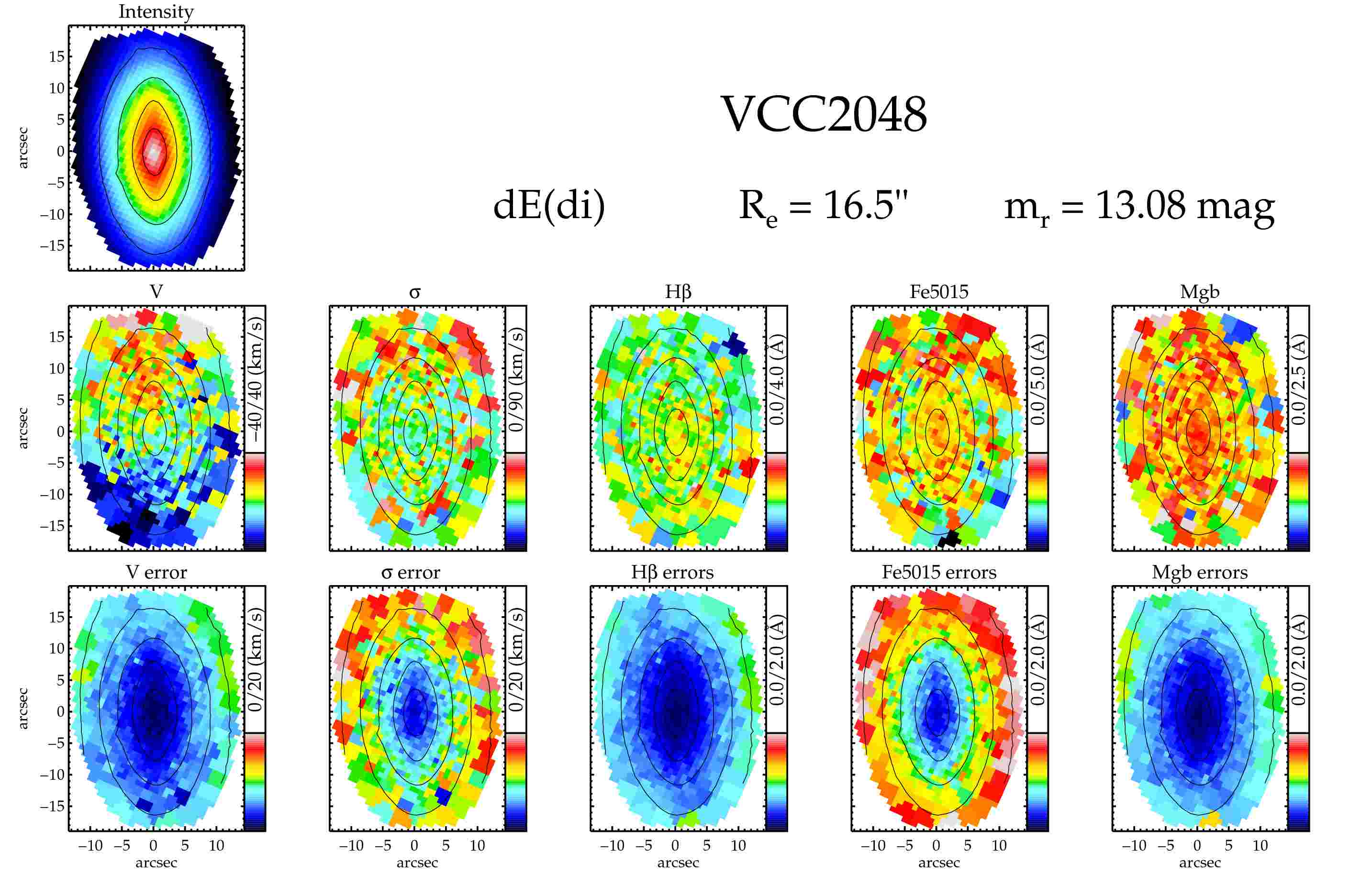}
\end{center}
\caption{Continued.}
\label{ppxf-maps2048}
\end{figure*}

\addtocounter{figure}{-1}
\addtocounter{subfigure}{0}

\clearpage
\begin{figure*}
\begin{center}
  \includegraphics[width=1.25\textwidth,angle=90]{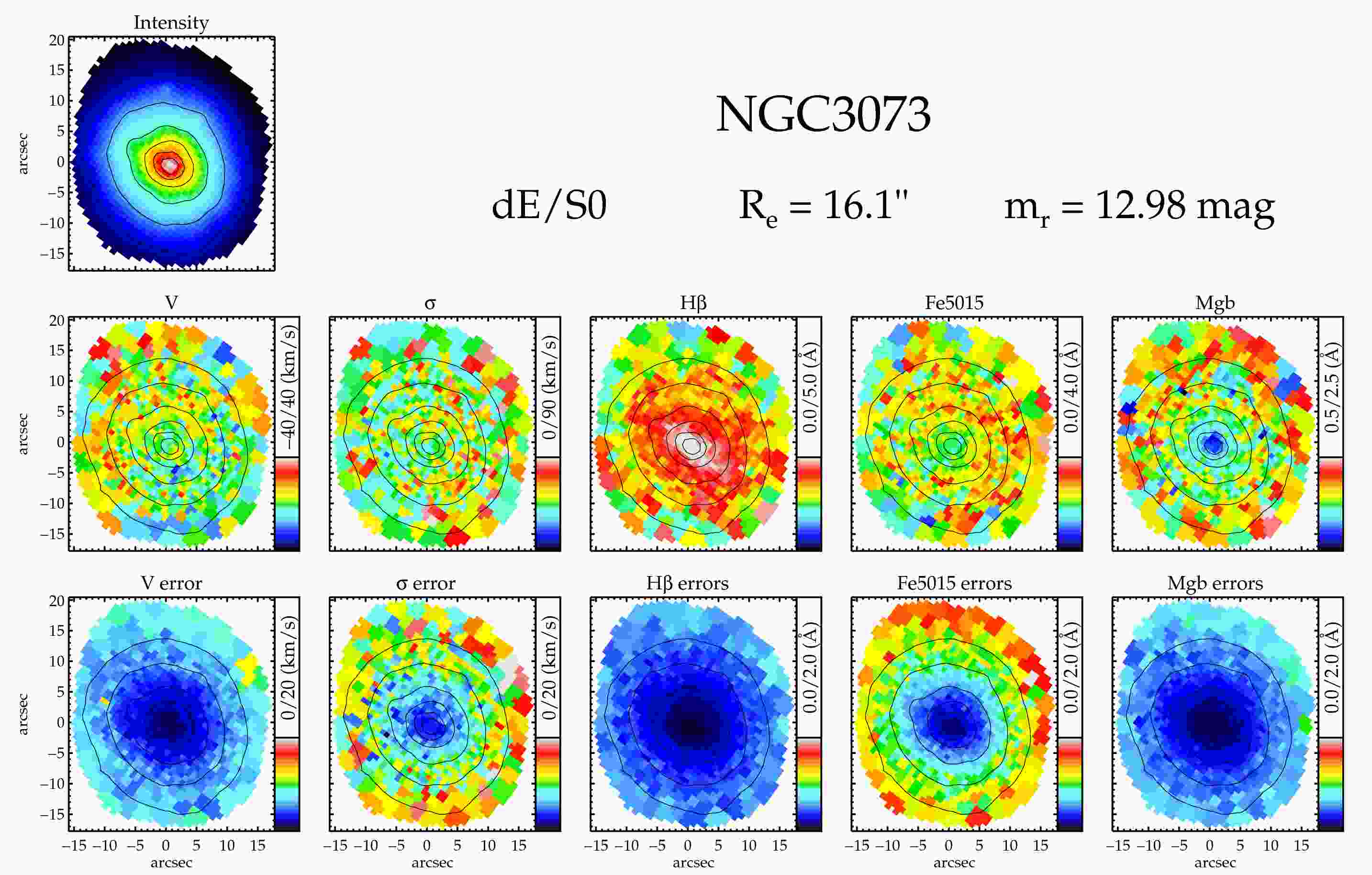}
\end{center}
\caption{Continued.}
\label{ppxf-maps3073}
\end{figure*}

\addtocounter{figure}{-1}
\addtocounter{subfigure}{0}

\clearpage
\begin{figure*}
\begin{center}
  \includegraphics[width=1.25\textwidth,angle=90]{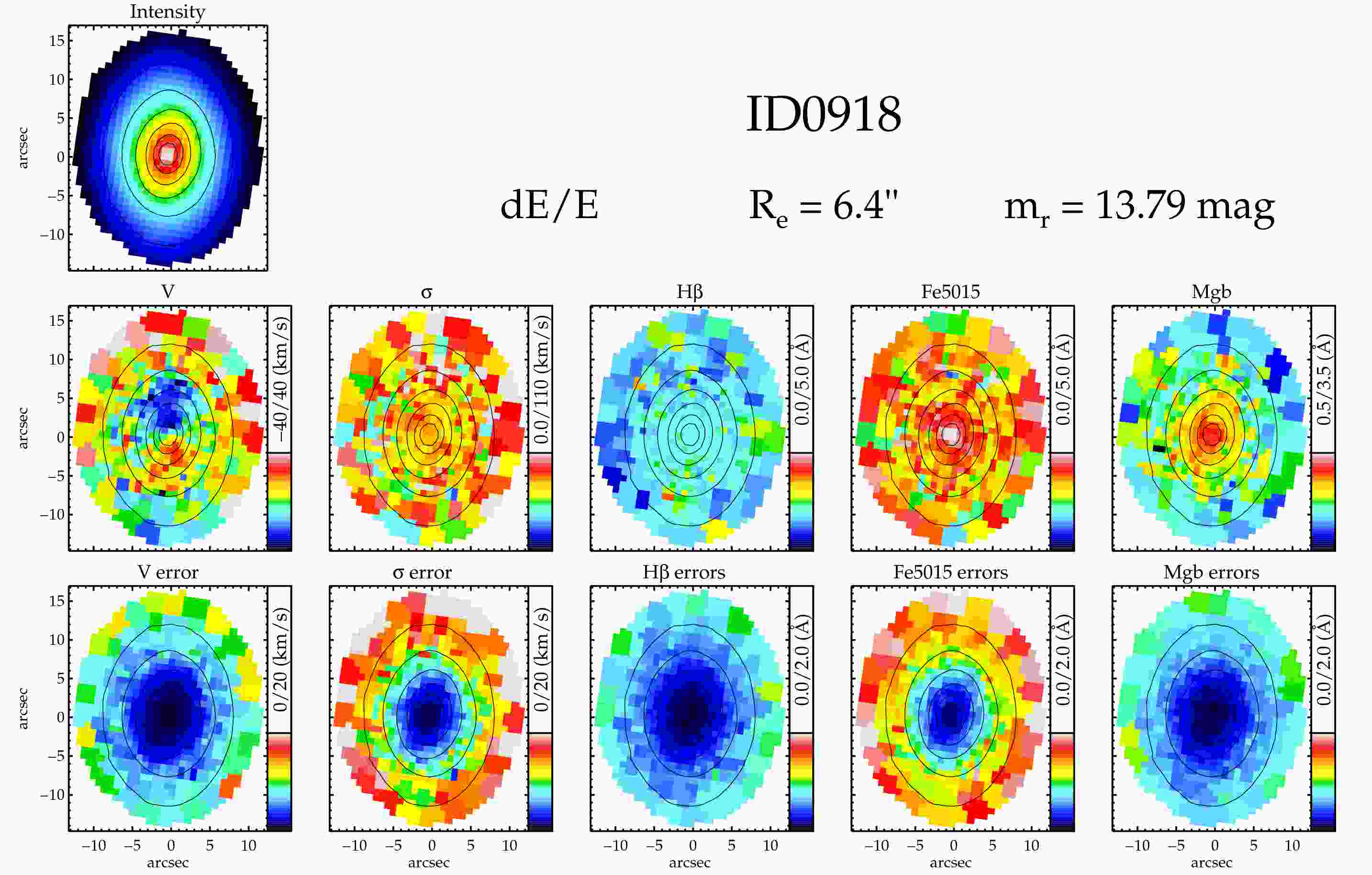}
\end{center}
\caption{Continued.}
\label{ppxf-maps0650}
\end{figure*}

\addtocounter{figure}{-1}
\addtocounter{subfigure}{0}

\clearpage
\begin{figure*}
\begin{center}
  \includegraphics[width=1.25\textwidth,angle=90]{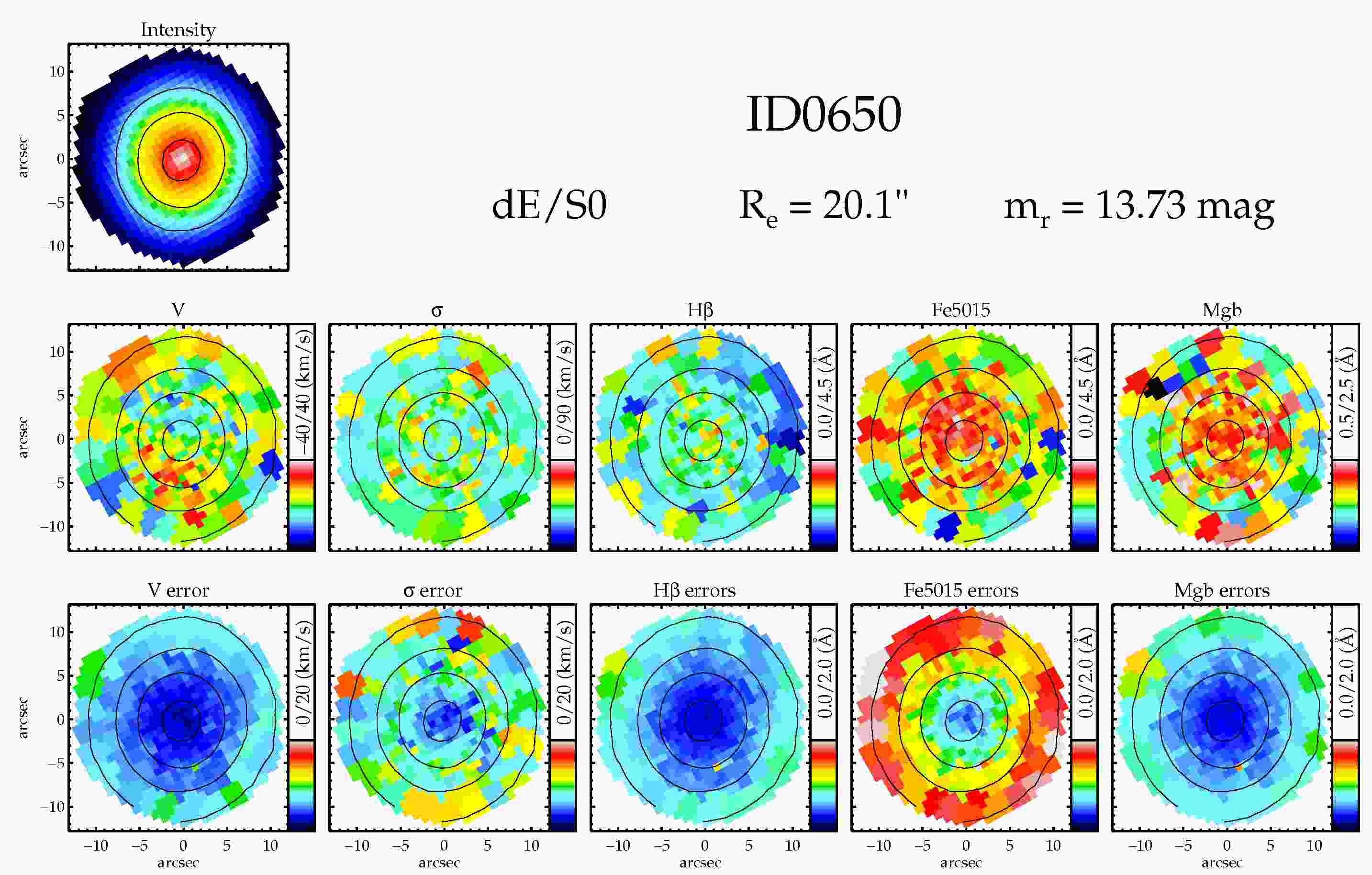}
\end{center}
\caption{Continued.}
\label{ppxf-maps0918}
\end{figure*}

\clearpage   

\begin{figure*}
\begin{center}
  \includegraphics[width=0.58\textwidth]{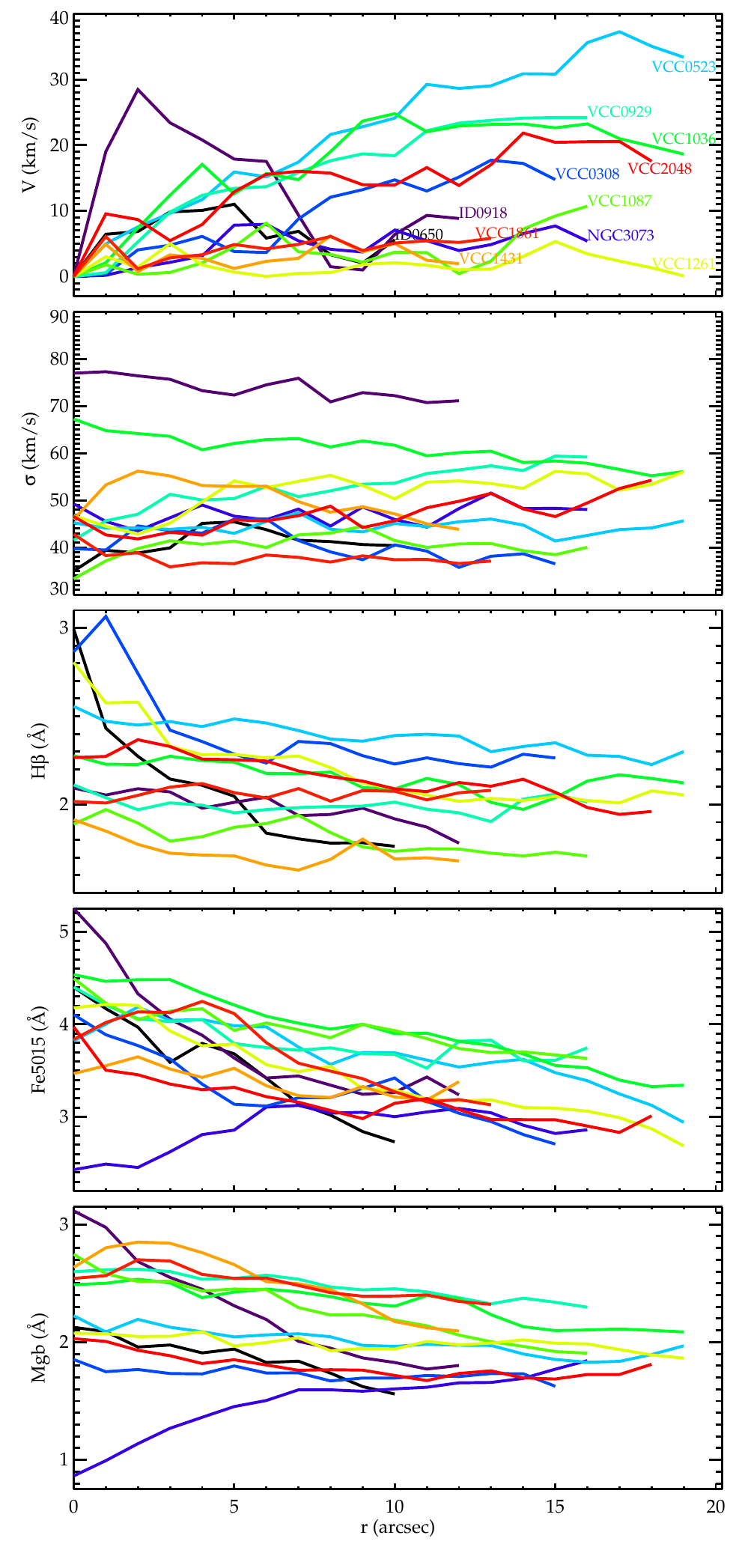}
\end{center}
\caption{One-dimensional (1D) profiles extracted from our maps, shown together to illustrate differences among our galaxies and to facilitate the comparison between our and literature data (the vast majority of the latter available in 1D). The errors, not shown here for clarity, are listed in Table~\ref{1Dprofiles-1}. In the H$\beta$ panel the NGC3073 profile (with values in the 5.26--3.02 range) is not included since the y-axis range is restricted to emphasize other galaxies' profile shapes.}
\label{all-profiles}
\end{figure*}

\begin{table*}
\caption{One-dimensional profiles of the stellar velocity $V$, velocity disperson $\sigma$, and H$\beta$, Fe5015, and Mg$b$ lines, and their errors, as a function of elliptical radius. The profiles were extracted in steps of 1'' using the \textit{kinemetry} software of \protect\cite{krajnovic:2006}.}%\\
\centering
\begin{tabular}{rrrrrrrrrrr}
\hline
\hline
r  & V  & eV & $\sigma$&e$\sigma$&H$\beta$&eH$\beta$&Fe5015  &eFe5015 & Mgb & eMgb\\
('') &(km/s)&(km/s)& (km/s)    & (km/s)    & (\AA)    & (\AA)     &(\AA)     &(\AA)     &(\AA)  & (\AA) \\
(1)&(2) &(3) &(4)      &(5)      &(6)     &(7)      &(8)     &(9)     &(10) &(11) \\
\hline
\hline
\multicolumn{11}{c}{VCC\,0308}\\
\hline
0.0 & 0.00 & 1.7 & 39.7 & 5.24 & 2.86 & 0.16 & 4.10 & 0.31 & 1.85 & 0.20 \\
1.0 & 0.36 & 1.9 & 39.4 & 5.05 & 3.06 & 0.21 & 3.88 & 0.43 & 1.74 & 0.23 \\
2.0 & 4.05 & 2.4 & 44.5 & 6.13 & 2.74 & 0.25 & 3.77 & 0.55 & 1.76 & 0.28 \\
3.0 & 4.78 & 3.0 & 43.5 & 7.17 & 2.42 & 0.33 & 3.62 & 0.70 & 1.73 & 0.35 \\
4.0 & 6.11 & 3.6 & 43.0 & 7.68 & 2.35 & 0.39 & 3.35 & 0.83 & 1.73 & 0.42 \\
5.0 & 3.82 & 4.2 & 46.0 & 8.89 & 2.28 & 0.45 & 3.13 & 0.96 & 1.79 & 0.48 \\
6.0 & 3.65 & 4.7 & 46.0 & 9.56 & 2.23 & 0.52 & 3.11 & 1.08 & 1.73 & 0.55 \\
7.0 & 8.76 & 5.0 & 41.4 & 9.03 & 2.35 & 0.54 & 3.20 & 1.14 & 1.73 & 0.57 \\
8.0 & 12.0 & 5.3 & 39.0 & 8.59 & 2.34 & 0.56 & 3.20 & 1.16 & 1.67 & 0.59 \\
9.0 & 13.2 & 5.5 & 37.3 & 8.61 & 2.27 & 0.59 & 3.30 & 1.21 & 1.69 & 0.62 \\
10. & 14.7 & 5.8 & 40.5 & 9.73 & 2.22 & 0.62 & 3.42 & 1.25 & 1.69 & 0.64 \\
11. & 13.0 & 6.1 & 39.2 & 9.37 & 2.26 & 0.66 & 3.16 & 1.35 & 1.71 & 0.68 \\
12. & 15.1 & 6.2 & 35.7 & 7.85 & 2.23 & 0.69 & 3.04 & 1.43 & 1.70 & 0.72 \\
13. & 17.7 & 6.6 & 38.1 & 8.86 & 2.21 & 0.73 & 2.95 & 1.50 & 1.73 & 0.76 \\
14. & 17.2 & 7.2 & 38.6 & 9.74 & 2.28 & 0.78 & 2.81 & 1.58 & 1.73 & 0.81 \\
15. & 14.7 & 7.2 & 36.5 & 8.40 & 2.26 & 0.78 & 2.70 & 1.53 & 1.62 & 0.80 \\
\hline

\multicolumn{11}{c}{VCC\,0523}\\
\hline
0.0 & 0.00 & 1.5 & 45.2 & 3.76 & 2.55 & 0.19 & 3.82 & 0.40 & 2.22 & 0.18 \\
1.0 & 4.98 & 1.6 & 44.1 & 3.50 & 2.47 & 0.18 & 4.00 & 0.40 & 2.08 & 0.20 \\
2.0 & 7.55 & 1.9 & 44.1 & 4.24 & 2.44 & 0.20 & 4.18 & 0.40 & 2.19 & 0.22 \\
3.0 & 9.63 & 2.1 & 43.8 & 4.77 & 2.46 & 0.22 & 4.04 & 0.47 & 2.12 & 0.24 \\
4.0 & 11.7 & 2.4 & 44.2 & 4.98 & 2.44 & 0.24 & 4.05 & 0.52 & 2.08 & 0.26 \\
5.0 & 15.9 & 2.7 & 42.9 & 5.57 & 2.48 & 0.28 & 3.98 & 0.59 & 2.04 & 0.29 \\
6.0 & 15.1 & 3.0 & 45.5 & 6.30 & 2.46 & 0.31 & 3.97 & 0.66 & 2.06 & 0.34 \\
7.0 & 17.4 & 3.3 & 47.3 & 6.73 & 2.41 & 0.34 & 3.75 & 0.73 & 2.07 & 0.37 \\
8.0 & 21.6 & 3.5 & 43.8 & 7.07 & 2.37 & 0.37 & 3.56 & 0.79 & 2.04 & 0.41 \\
9.0 & 22.8 & 3.8 & 43.2 & 7.45 & 2.35 & 0.40 & 3.69 & 0.87 & 1.97 & 0.44 \\
10. & 24.1 & 4.1 & 45.0 & 7.98 & 2.39 & 0.42 & 3.69 & 0.92 & 1.96 & 0.47 \\
11. & 29.2 & 4.3 & 44.4 & 8.33 & 2.39 & 0.44 & 3.61 & 0.96 & 1.98 & 0.48 \\
12. & 28.6 & 4.5 & 45.4 & 8.39 & 2.38 & 0.46 & 3.53 & 0.99 & 1.97 & 0.50 \\
13. & 29.0 & 4.8 & 46.0 & 8.82 & 2.29 & 0.49 & 3.58 & 1.05 & 1.97 & 0.53 \\
14. & 30.8 & 5.0 & 44.7 & 9.07 & 2.32 & 0.50 & 3.62 & 1.08 & 1.90 & 0.55 \\
15. & 30.8 & 5.1 & 41.3 & 8.66 & 2.34 & 0.51 & 3.47 & 1.12 & 1.85 & 0.56 \\
16. & 35.5 & 5.4 & 42.5 & 9.19 & 2.27 & 0.54 & 3.39 & 1.17 & 1.82 & 0.60 \\
17. & 37.2 & 5.6 & 43.8 & 9.73 & 2.27 & 0.57 & 3.24 & 1.23 & 1.83 & 0.62 \\
18. & 35.0 & 5.9 & 44.1 & 9.93 & 2.22 & 0.59 & 3.12 & 1.30 & 1.89 & 0.65 \\
19. & 33.3 & 6.2 & 45.6 & 10.6 & 2.29 & 0.62 & 2.94 & 1.37 & 1.96 & 0.68 \\
\hline

\label{1Dprofiles-1} 
\end{tabular}
\end{table*}

\newpage

\addtocounter{table}{-1}

\begin{table*}
\caption{Continued.} 
\centering
\begin{tabular}{rrrrrrrrrrr}
\hline
\hline
r  & V  & eV & $\sigma$&e$\sigma$&H$\beta$&eH$\beta$&Fe5015  &eFe5015 & Mgb & eMgb\\
('') &(km/s)&(km/s)& (km/s)    & (km/s)    & (\AA)    & (\AA)     &(\AA)     &(\AA)     &(\AA)  & (\AA) \\
(1)&(2) &(3) &(4)      &(5)      &(6)     &(7)      &(8)     &(9)     &(10) &(11) \\
\hline
\hline
\multicolumn{11}{c}{VCC\,0929}\\
\hline
0.0 & 0.00 & 1.0 & 41.5 & 3.05 & 2.11 & 0.12 & 4.40 & 0.28 & 2.59 & 0.13 \\
1.0 & 0.59 & 1.2 & 45.6 & 3.61 & 2.03 & 0.17 & 4.20 & 0.36 & 2.61 & 0.17 \\
2.0 & 5.29 & 1.4 & 47.0 & 4.26 & 1.97 & 0.18 & 4.06 & 0.40 & 2.62 & 0.20 \\
3.0 & 9.81 & 1.7 & 51.2 & 4.35 & 2.00 & 0.20 & 4.02 & 0.44 & 2.60 & 0.22 \\
4.0 & 12.3 & 1.8 & 50.0 & 4.68 & 1.99 & 0.22 & 4.05 & 0.48 & 2.53 & 0.24 \\
5.0 & 13.4 & 2.0 & 50.3 & 5.16 & 1.95 & 0.24 & 3.79 & 0.51 & 2.54 & 0.26 \\
6.0 & 13.6 & 2.3 & 52.9 & 5.80 & 1.97 & 0.26 & 3.74 & 0.59 & 2.56 & 0.29 \\
7.0 & 15.8 & 2.7 & 50.7 & 6.35 & 1.98 & 0.29 & 3.71 & 0.65 & 2.53 & 0.32 \\
8.0 & 17.6 & 3.0 & 52.0 & 7.11 & 1.98 & 0.33 & 3.74 & 0.72 & 2.46 & 0.36 \\
9.0 & 18.6 & 3.3 & 53.4 & 7.58 & 1.99 & 0.37 & 3.68 & 0.80 & 2.44 & 0.40 \\
10. & 18.4 & 3.8 & 53.6 & 8.44 & 2.01 & 0.42 & 3.67 & 0.91 & 2.45 & 0.44 \\
11. & 22.2 & 4.1 & 55.7 & 8.88 & 1.97 & 0.45 & 3.52 & 0.98 & 2.42 & 0.48 \\
12. & 23.3 & 4.6 & 56.4 & 9.82 & 1.95 & 0.48 & 3.81 & 1.06 & 2.37 & 0.52 \\
13. & 23.7 & 4.8 & 57.3 & 10.2 & 1.90 & 0.51 & 3.82 & 1.11 & 2.32 & 0.56 \\
14. & 24.1 & 5.1 & 56.3 & 10.1 & 2.03 & 0.54 & 3.60 & 1.15 & 2.37 & 0.57 \\
15. & 24.2 & 5.5 & 59.4 & 10.8 & 2.05 & 0.58 & 3.61 & 1.25 & 2.33 & 0.62 \\
16. & 24.2 & 5.9 & 59.1 & 11.5 & 2.01 & 0.61 & 3.74 & 1.32 & 2.29 & 0.66 \\
\hline

\multicolumn{11}{c}{VCC\,1036}\\
\hline
0.0 & 0.00 & 1.2 & 67.2 & 2.81 & 2.27 & 0.18 & 4.53 & 0.37 & 2.48 & 0.19 \\
1.0 & 2.11 & 1.4 & 64.8 & 3.09 & 2.22 & 0.18 & 4.46 & 0.37 & 2.50 & 0.19 \\
2.0 & 7.40 & 1.5 & 64.1 & 3.17 & 2.22 & 0.17 & 4.48 & 0.36 & 2.53 & 0.19 \\
3.0 & 12.3 & 1.7 & 63.5 & 3.67 & 2.27 & 0.19 & 4.48 & 0.41 & 2.50 & 0.21 \\
4.0 & 17.0 & 1.9 & 60.7 & 3.95 & 2.24 & 0.22 & 4.33 & 0.46 & 2.37 & 0.23 \\
5.0 & 12.5 & 2.0 & 62.0 & 4.24 & 2.23 & 0.23 & 4.20 & 0.50 & 2.42 & 0.25 \\
6.0 & 15.6 & 2.2 & 62.9 & 4.59 & 2.17 & 0.24 & 4.08 & 0.53 & 2.45 & 0.27 \\
7.0 & 14.7 & 2.5 & 63.1 & 5.14 & 2.17 & 0.28 & 4.01 & 0.58 & 2.42 & 0.29 \\
8.0 & 19.0 & 2.8 & 61.3 & 5.86 & 2.18 & 0.30 & 3.94 & 0.65 & 2.38 & 0.33 \\
9.0 & 23.6 & 3.1 & 62.6 & 6.32 & 2.09 & 0.33 & 4.00 & 0.71 & 2.33 & 0.36 \\
10. & 24.7 & 3.5 & 61.7 & 6.89 & 2.08 & 0.37 & 3.89 & 0.78 & 2.30 & 0.40 \\
11. & 22.0 & 3.8 & 59.4 & 7.65 & 2.14 & 0.41 & 3.90 & 0.86 & 2.39 & 0.43 \\
12. & 22.9 & 4.2 & 60.1 & 8.32 & 2.11 & 0.44 & 3.81 & 0.94 & 2.36 & 0.46 \\
13. & 23.1 & 4.4 & 60.4 & 8.40 & 2.01 & 0.46 & 3.77 & 0.97 & 2.23 & 0.49 \\
14. & 23.2 & 4.7 & 58.0 & 9.09 & 1.97 & 0.48 & 3.68 & 1.02 & 2.13 & 0.51 \\
15. & 22.6 & 5.1 & 58.3 & 9.87 & 2.03 & 0.52 & 3.55 & 1.13 & 2.09 & 0.56 \\
16. & 23.2 & 5.3 & 57.8 & 10.2 & 2.13 & 0.56 & 3.53 & 1.20 & 2.10 & 0.59 \\
17. & 20.9 & 5.4 & 56.5 & 10.4 & 2.16 & 0.59 & 3.39 & 1.24 & 2.11 & 0.62 \\
18. & 19.8 & 5.7 & 55.2 & 10.5 & 2.14 & 0.62 & 3.32 & 1.32 & 2.10 & 0.65 \\
19. & 18.6 & 6.0 & 56.1 & 10.8 & 2.12 & 0.66 & 3.34 & 1.40 & 2.08 & 0.69 \\
\hline

\label{1Dprofiles-2} 
\end{tabular}
\end{table*}

\newpage

\addtocounter{table}{-1}

\begin{table*}
\caption{Continued.} 
\centering
\begin{tabular}{rrrrrrrrrrr}
\hline
\hline
r  & V  & eV & $\sigma$&e$\sigma$&H$\beta$&eH$\beta$&Fe5015  &eFe5015 & Mgb & eMgb\\
('') &(km/s)&(km/s)& (km/s)    & (km/s)    & (\AA)    & (\AA)     &(\AA)     &(\AA)     &(\AA)  & (\AA) \\
(1)&(2) &(3) &(4)      &(5)      &(6)     &(7)      &(8)     &(9)     &(10) &(11) \\
\hline
\hline
\multicolumn{11}{c}{VCC\,1087}\\
\hline
0.0 & 0.00 & 2.0 & 33.3 & 3.69 & 1.88 & 0.17 & 4.49 & 0.38 & 2.74 & 0.16 \\
1.0 & 1.78 & 2.2 & 37.1 & 4.63 & 1.97 & 0.22 & 4.22 & 0.46 & 2.58 & 0.23 \\
2.0 & 0.34 & 2.6 & 39.7 & 5.24 & 1.89 & 0.26 & 4.05 & 0.59 & 2.51 & 0.29 \\
3.0 & 0.62 & 2.7 & 41.4 & 6.11 & 1.79 & 0.29 & 4.14 & 0.63 & 2.51 & 0.31 \\
4.0 & 2.07 & 3.0 & 40.6 & 6.14 & 1.81 & 0.32 & 4.16 & 0.71 & 2.43 & 0.35 \\
5.0 & 4.53 & 3.3 & 41.3 & 6.88 & 1.87 & 0.36 & 3.93 & 0.80 & 2.45 & 0.39 \\
6.0 & 8.13 & 3.6 & 39.9 & 6.93 & 1.89 & 0.39 & 4.01 & 0.88 & 2.44 & 0.44 \\
7.0 & 3.76 & 4.1 & 42.6 & 7.74 & 1.93 & 0.43 & 3.94 & 0.95 & 2.29 & 0.48 \\
8.0 & 3.37 & 4.2 & 43.0 & 8.02 & 1.84 & 0.43 & 3.85 & 0.94 & 2.23 & 0.48 \\
9.0 & 2.20 & 4.5 & 44.5 & 8.30 & 1.76 & 0.45 & 3.99 & 0.98 & 2.23 & 0.49 \\
10. & 3.72 & 4.8 & 41.4 & 8.18 & 1.73 & 0.48 & 3.93 & 1.05 & 2.18 & 0.53 \\
11. & 3.63 & 4.9 & 40.0 & 8.06 & 1.74 & 0.48 & 3.84 & 1.09 & 2.13 & 0.55 \\
12. & 0.44 & 5.2 & 40.7 & 8.68 & 1.74 & 0.51 & 3.73 & 1.13 & 2.06 & 0.57 \\
13. & 2.37 & 5.4 & 40.8 & 8.74 & 1.72 & 0.53 & 3.69 & 1.15 & 2.00 & 0.59 \\
14. & 7.22 & 5.5 & 39.3 & 8.07 & 1.70 & 0.53 & 3.70 & 1.17 & 1.96 & 0.60 \\
15. & 9.22 & 5.6 & 38.4 & 7.90 & 1.72 & 0.53 & 3.66 & 1.20 & 1.91 & 0.61 \\
16. & 10.7 & 6.0 & 40.0 & 8.03 & 1.70 & 0.56 & 3.63 & 1.26 & 1.90 & 0.64 \\
\hline

\multicolumn{11}{c}{VCC\,1261}\\
\hline
0.0 & 0.00 & 1.3 & 46.7 & 3.99 & 2.80 & 0.15 & 4.17 & 0.34 & 2.07 & 0.18 \\
1.0 & 3.02 & 1.3 & 44.5 & 4.02 & 2.57 & 0.15 & 4.21 & 0.33 & 2.07 & 0.17 \\
2.0 & 1.42 & 1.7 & 42.8 & 4.43 & 2.57 & 0.19 & 4.20 & 0.40 & 2.04 & 0.20 \\
3.0 & 5.00 & 1.9 & 45.1 & 4.98 & 2.32 & 0.21 & 3.92 & 0.45 & 2.05 & 0.23 \\
4.0 & 1.76 & 2.1 & 49.6 & 5.38 & 2.28 & 0.22 & 3.76 & 0.50 & 2.08 & 0.24 \\
5.0 & 0.68 & 2.4 & 54.1 & 5.74 & 2.28 & 0.25 & 3.78 & 0.54 & 1.96 & 0.27 \\
6.0 & 0.03 & 2.8 & 52.7 & 6.48 & 2.26 & 0.28 & 3.56 & 0.62 & 1.99 & 0.31 \\
7.0 & 0.48 & 3.1 & 54.0 & 7.04 & 2.27 & 0.32 & 3.48 & 0.69 & 2.03 & 0.34 \\
8.0 & 0.61 & 3.6 & 55.3 & 7.82 & 2.20 & 0.36 & 3.54 & 0.76 & 1.92 & 0.38 \\
9.0 & 1.82 & 3.9 & 53.1 & 8.76 & 2.12 & 0.38 & 3.30 & 0.84 & 1.94 & 0.42 \\
10. & 2.12 & 4.2 & 50.3 & 9.00 & 2.09 & 0.40 & 3.29 & 0.90 & 1.94 & 0.44 \\
11. & 1.72 & 4.4 & 53.8 & 9.27 & 2.05 & 0.43 & 3.19 & 0.95 & 2.00 & 0.47 \\
12. & 1.04 & 4.6 & 54.1 & 9.67 & 2.01 & 0.46 & 3.17 & 1.02 & 1.97 & 0.49 \\
13. & 1.14 & 5.0 & 53.5 & 10.0 & 2.03 & 0.48 & 3.18 & 1.06 & 1.99 & 0.52 \\
14. & 3.06 & 5.5 & 52.5 & 10.1 & 2.02 & 0.49 & 3.10 & 1.11 & 2.01 & 0.54 \\
15. & 5.34 & 5.7 & 56.1 & 10.6 & 2.04 & 0.51 & 3.09 & 1.17 & 1.99 & 0.55 \\
16. & 3.45 & 6.1 & 55.6 & 10.7 & 2.02 & 0.55 & 3.06 & 1.23 & 1.98 & 0.59 \\
17. & 2.36 & 6.5 & 52.1 & 10.5 & 2.00 & 0.57 & 2.99 & 1.28 & 1.93 & 0.61 \\
18. & 1.33 & 6.1 & 53.4 & 10.7 & 2.07 & 0.56 & 2.87 & 1.27 & 1.89 & 0.60 \\
19. & 0.10 & 6.2 & 56.0 & 11.6 & 2.05 & 0.56 & 2.68 & 1.28 & 1.86 & 0.61 \\
\hline

\multicolumn{11}{c}{VCC\,1431}\\
\hline
0.0 & 0.00 & 1.0 & 45.9 & 2.44 & 1.91 & 0.10 & 3.46 & 0.22 & 2.63 & 0.10 \\
1.0 & 4.92 & 1.2 & 53.2 & 3.11 & 1.84 & 0.14 & 3.55 & 0.31 & 2.80 & 0.15 \\
2.0 & 0.75 & 1.6 & 56.2 & 3.79 & 1.77 & 0.19 & 3.64 & 0.40 & 2.85 & 0.20 \\
3.0 & 3.31 & 2.2 & 55.1 & 4.90 & 1.72 & 0.24 & 3.51 & 0.52 & 2.84 & 0.26 \\
4.0 & 2.78 & 2.6 & 53.1 & 6.07 & 1.71 & 0.28 & 3.42 & 0.60 & 2.76 & 0.30 \\
5.0 & 1.23 & 3.1 & 52.9 & 6.81 & 1.70 & 0.32 & 3.52 & 0.69 & 2.65 & 0.33 \\
6.0 & 2.32 & 3.9 & 52.9 & 8.07 & 1.65 & 0.40 & 3.33 & 0.84 & 2.51 & 0.41 \\
7.0 & 2.78 & 4.7 & 49.7 & 8.81 & 1.62 & 0.46 & 3.23 & 0.98 & 2.49 & 0.47 \\
8.0 & 6.16 & 5.4 & 47.4 & 9.32 & 1.68 & 0.53 & 3.20 & 1.11 & 2.44 & 0.55 \\
9.0 & 3.77 & 5.7 & 48.6 & 10.2 & 1.80 & 0.55 & 3.32 & 1.15 & 2.32 & 0.58 \\
10. & 5.01 & 6.3 & 47.1 & 10.5 & 1.69 & 0.60 & 3.21 & 1.25 & 2.17 & 0.62 \\
11. & 2.47 & 7.0 & 45.0 & 10.4 & 1.69 & 0.67 & 3.18 & 1.37 & 2.12 & 0.67 \\
12. & 1.95 & 7.6 & 43.8 & 10.5 & 1.68 & 0.75 & 3.38 & 1.54 & 2.09 & 0.75 \\
\hline

\label{1Dprofiles-3} 
\end{tabular}
\end{table*}

\newpage

\addtocounter{table}{-1}

\begin{table*}
\caption{Continued.} %\\
\centering
\begin{tabular}{rrrrrrrrrrr}
\hline
\hline
r  & V  & eV & $\sigma$&e$\sigma$&H$\beta$&eH$\beta$&Fe5015  &eFe5015 & Mgb & eMgb\\
('') &(km/s)&(km/s)& (km/s)    & (km/s)    & (\AA)    & (\AA)     &(\AA)     &(\AA)     &(\AA)  & (\AA) \\
(1)&(2) &(3) &(4)      &(5)      &(6)     &(7)      &(8)     &(9)     &(10) &(11) \\
\hline
\hline
\multicolumn{11}{c}{VCC\,1861}\\
\hline
0.0 & 0.00 & 1.6 & 42.8 & 4.02 & 2.01 & 0.19 & 3.83 & 0.38 & 2.54 & 0.20 \\
1.0 & 6.08 & 2.4 & 38.2 & 4.24 & 2.00 & 0.24 & 4.01 & 0.52 & 2.56 & 0.26 \\
2.0 & 1.18 & 2.7 & 38.8 & 4.74 & 2.05 & 0.28 & 4.13 & 0.61 & 2.69 & 0.31 \\
3.0 & 2.84 & 3.0 & 35.8 & 4.95 & 2.09 & 0.33 & 4.12 & 0.73 & 2.68 & 0.35 \\
4.0 & 3.32 & 3.2 & 36.7 & 5.54 & 2.11 & 0.35 & 4.24 & 0.76 & 2.57 & 0.39 \\
5.0 & 4.88 & 3.6 & 36.5 & 5.81 & 2.06 & 0.40 & 4.11 & 0.85 & 2.54 & 0.44 \\
6.0 & 4.23 & 4.1 & 38.3 & 6.49 & 2.03 & 0.44 & 3.80 & 0.96 & 2.54 & 0.48 \\
7.0 & 4.90 & 4.3 & 37.9 & 6.68 & 2.09 & 0.46 & 3.58 & 0.99 & 2.48 & 0.50 \\
8.0 & 6.11 & 4.7 & 36.9 & 7.00 & 2.01 & 0.50 & 3.49 & 1.05 & 2.42 & 0.55 \\
9.0 & 3.96 & 4.9 & 38.2 & 7.41 & 2.07 & 0.50 & 3.41 & 1.12 & 2.39 & 0.57 \\
10. & 5.10 & 5.3 & 37.3 & 7.61 & 2.07 & 0.54 & 3.27 & 1.21 & 2.39 & 0.60 \\
11. & 5.44 & 5.7 & 37.4 & 7.70 & 2.02 & 0.57 & 3.16 & 1.27 & 2.40 & 0.62 \\
12. & 5.19 & 5.6 & 36.5 & 7.50 & 2.06 & 0.57 & 3.18 & 1.26 & 2.34 & 0.63 \\
13. & 5.81 & 5.9 & 37.1 & 7.96 & 2.08 & 0.60 & 3.13 & 1.30 & 2.32 & 0.65 \\
\hline

\multicolumn{11}{c}{VCC\,2048}\\
\hline
0.0 & 0.00 & 1.0 & 46.5 & 2.15 & 2.26 & 0.10 & 3.97 & 0.25 & 2.02 & 0.12 \\
1.0 & 9.55 & 1.2 & 42.6 & 2.85 & 2.27 & 0.13 & 3.50 & 0.28 & 2.00 & 0.14 \\
2.0 & 8.67 & 1.5 & 41.8 & 3.39 & 2.36 & 0.15 & 3.45 & 0.33 & 1.93 & 0.16 \\
3.0 & 5.48 & 1.8 & 43.2 & 3.92 & 2.32 & 0.18 & 3.35 & 0.39 & 1.88 & 0.19 \\
4.0 & 7.91 & 2.1 & 42.6 & 4.46 & 2.25 & 0.20 & 3.29 & 0.45 & 1.81 & 0.22 \\
5.0 & 12.8 & 2.3 & 45.8 & 5.10 & 2.25 & 0.23 & 3.31 & 0.49 & 1.85 & 0.24 \\
6.0 & 15.5 & 2.7 & 45.6 & 5.77 & 2.24 & 0.25 & 3.21 & 0.55 & 1.80 & 0.27 \\
7.0 & 16.0 & 2.9 & 46.7 & 6.13 & 2.19 & 0.28 & 3.15 & 0.62 & 1.76 & 0.31 \\
8.0 & 15.7 & 3.3 & 48.7 & 6.90 & 2.15 & 0.32 & 3.06 & 0.71 & 1.76 & 0.34 \\
9.0 & 13.9 & 3.9 & 44.2 & 7.25 & 2.13 & 0.35 & 2.98 & 0.77 & 1.76 & 0.38 \\
10. & 13.9 & 4.4 & 45.6 & 8.19 & 2.08 & 0.41 & 3.14 & 0.89 & 1.71 & 0.44 \\
11. & 16.5 & 4.8 & 48.4 & 9.06 & 2.07 & 0.45 & 3.19 & 1.00 & 1.67 & 0.49 \\
12. & 13.8 & 5.2 & 49.8 & 9.31 & 2.12 & 0.50 & 3.07 & 1.07 & 1.73 & 0.53 \\
13. & 17.0 & 5.5 & 51.4 & 10.2 & 2.10 & 0.53 & 2.97 & 1.12 & 1.75 & 0.55 \\
14. & 21.8 & 5.7 & 48.1 & 9.81 & 2.14 & 0.55 & 2.97 & 1.18 & 1.69 & 0.58 \\
15. & 20.4 & 5.9 & 46.5 & 9.89 & 2.06 & 0.58 & 2.96 & 1.25 & 1.68 & 0.62 \\
16. & 20.5 & 6.1 & 49.4 & 10.7 & 1.98 & 0.62 & 2.90 & 1.31 & 1.72 & 0.66 \\
17. & 20.5 & 6.6 & 52.5 & 12.1 & 1.94 & 0.67 & 2.83 & 1.42 & 1.72 & 0.71 \\
18. & 17.5 & 6.9 & 54.2 & 13.2 & 1.96 & 0.73 & 3.01 & 1.54 & 1.81 & 0.77 \\
\hline

\multicolumn{11}{c}{NGC\,3073}\\
\hline
0.0 & 0.00 & 1.3 & 49.2 & 3.36 & 5.26 & 0.08 & 2.43 & 0.20 & 0.86 & 0.09 \\
1.0 & 0.19 & 1.5 & 45.5 & 3.79 & 5.12 & 0.09 & 2.49 & 0.20 & 0.99 & 0.10 \\
2.0 & 1.33 & 1.7 & 43.5 & 4.18 & 4.88 & 0.10 & 2.45 & 0.24 & 1.13 & 0.12 \\
3.0 & 2.17 & 2.1 & 46.2 & 4.81 & 4.66 & 0.14 & 2.62 & 0.30 & 1.26 & 0.15 \\
4.0 & 3.13 & 2.6 & 49.0 & 5.69 & 4.43 & 0.18 & 2.80 & 0.39 & 1.36 & 0.20 \\
5.0 & 7.82 & 2.8 & 46.6 & 6.26 & 4.25 & 0.21 & 2.85 & 0.46 & 1.45 & 0.24 \\
6.0 & 7.95 & 3.3 & 45.8 & 6.46 & 4.09 & 0.24 & 3.10 & 0.53 & 1.50 & 0.27 \\
7.0 & 5.44 & 3.6 & 48.1 & 7.69 & 3.91 & 0.28 & 3.12 & 0.61 & 1.59 & 0.31 \\
8.0 & 4.15 & 4.1 & 44.5 & 7.24 & 3.76 & 0.31 & 3.04 & 0.70 & 1.59 & 0.36 \\
9.0 & 3.74 & 4.7 & 48.6 & 8.79 & 3.76 & 0.35 & 3.05 & 0.81 & 1.58 & 0.42 \\
10. & 7.07 & 5.5 & 45.9 & 9.47 & 3.65 & 0.41 & 3.00 & 0.93 & 1.60 & 0.47 \\
11. & 5.46 & 5.8 & 44.3 & 9.26 & 3.59 & 0.43 & 3.05 & 0.98 & 1.61 & 0.50 \\
12. & 3.96 & 5.9 & 48.2 & 9.95 & 3.37 & 0.44 & 3.09 & 1.01 & 1.65 & 0.50 \\
13. & 4.87 & 6.2 & 51.5 & 11.1 & 3.22 & 0.48 & 3.04 & 1.08 & 1.65 & 0.54 \\
14. & 6.54 & 6.4 & 48.2 & 10.7 & 3.15 & 0.52 & 2.91 & 1.15 & 1.69 & 0.59 \\
15. & 7.71 & 6.7 & 48.3 & 11.4 & 3.07 & 0.54 & 2.82 & 1.19 & 1.77 & 0.61 \\
16. & 5.39 & 7.1 & 48.0 & 11.8 & 3.08 & 0.54 & 2.86 & 1.24 & 1.84 & 0.62 \\
\hline

\label{1Dprofiles-4} 
\end{tabular}
\end{table*}

\newpage

\addtocounter{table}{-1}

\begin{table*}
\caption{Continued.} %\\
\centering
\begin{tabular}{rrrrrrrrrrr}
\hline
\hline
r  & V  & eV & $\sigma$&e$\sigma$&H$\beta$&eH$\beta$&Fe5015  &eFe5015 & Mgb & eMgb\\
('') &(km/s)&(km/s)& (km/s)    & (km/s)    & (\AA)    & (\AA)     &(\AA)     &(\AA)     &(\AA)  & (\AA) \\
(1)&(2) &(3) &(4)      &(5)      &(6)     &(7)      &(8)     &(9)     &(10) &(11) \\
\hline
\hline
\multicolumn{11}{c}{ID\,0650}\\
\hline
0.0 & 0.00 & 1.9 & 34.9 & 3.39 & 2.99 & 0.20 & 4.40 & 0.40 & 2.12 & 0.24 \\
1.0 & 6.43 & 2.4 & 39.4 & 4.53 & 2.43 & 0.26 & 4.16 & 0.55 & 2.08 & 0.28 \\
2.0 & 6.86 & 2.9 & 38.8 & 5.38 & 2.27 & 0.31 & 3.96 & 0.65 & 1.95 & 0.33 \\
3.0 & 9.83 & 3.4 & 39.9 & 5.93 & 2.14 & 0.36 & 3.58 & 0.76 & 1.97 & 0.38 \\
4.0 & 10.0 & 3.8 & 45.1 & 7.20 & 2.10 & 0.41 & 3.79 & 0.89 & 1.90 & 0.45 \\
5.0 & 11.0 & 4.3 & 45.4 & 8.11 & 2.04 & 0.47 & 3.68 & 1.02 & 1.94 & 0.51 \\
6.0 & 5.86 & 5.0 & 43.7 & 8.24 & 1.83 & 0.52 & 3.41 & 1.11 & 1.82 & 0.55 \\
7.0 & 6.88 & 5.3 & 41.5 & 8.49 & 1.80 & 0.57 & 3.14 & 1.21 & 1.83 & 0.60 \\
8.0 & 3.33 & 6.0 & 41.2 & 8.59 & 1.78 & 0.63 & 3.02 & 1.33 & 1.73 & 0.66 \\
9.0 & 2.06 & 6.6 & 40.6 & 9.02 & 1.78 & 0.70 & 2.84 & 1.46 & 1.62 & 0.73 \\
10. & 5.57 & 7.1 & 40.3 & 9.28 & 1.76 & 0.76 & 2.73 & 1.57 & 1.55 & 0.80 \\
\hline

\multicolumn{11}{c}{ID\,0918}\\
\hline
0.0 & 0.00 & 0.6 & 77.0 & 1.02 & 2.09 & 0.07 & 5.24 & 0.17 & 3.11 & 0.08 \\
1.0 & 19.0 & 0.8 & 77.3 & 1.50 & 2.05 & 0.10 & 4.87 & 0.21 & 2.97 & 0.11 \\
2.0 & 28.4 & 1.2 & 76.4 & 2.21 & 2.08 & 0.14 & 4.32 & 0.30 & 2.68 & 0.14 \\
3.0 & 23.4 & 1.6 & 75.7 & 3.07 & 2.07 & 0.18 & 4.05 & 0.39 & 2.54 & 0.19 \\
4.0 & 20.7 & 2.1 & 73.3 & 4.11 & 1.97 & 0.23 & 3.88 & 0.47 & 2.44 & 0.24 \\
5.0 & 17.8 & 2.8 & 72.3 & 5.41 & 2.01 & 0.29 & 3.63 & 0.63 & 2.30 & 0.31 \\
6.0 & 17.5 & 3.7 & 74.5 & 6.92 & 2.04 & 0.36 & 3.42 & 0.78 & 2.19 & 0.38 \\
7.0 & 9.35 & 4.7 & 75.9 & 8.47 & 1.93 & 0.44 & 3.44 & 0.93 & 2.00 & 0.46 \\
8.0 & 1.50 & 5.8 & 70.9 & 9.87 & 1.94 & 0.53 & 3.34 & 1.13 & 1.94 & 0.57 \\
9.0 & 0.99 & 6.5 & 72.8 & 11.7 & 1.97 & 0.56 & 3.24 & 1.20 & 1.86 & 0.60 \\
10. & 6.39 & 6.9 & 72.2 & 12.6 & 1.91 & 0.60 & 3.26 & 1.28 & 1.82 & 0.63 \\
11. & 9.30 & 7.2 & 70.7 & 13.5 & 1.87 & 0.65 & 3.43 & 1.35 & 1.77 & 0.67 \\
12. & 8.84 & 7.6 & 71.1 & 14.3 & 1.78 & 0.70 & 3.23 & 1.40 & 1.80 & 0.70 \\
\hline

\label{1Dprofiles-5} 
\end{tabular}
\end{table*}

\label{lastpage}

\end{document}